\documentclass[12pt]{iopart}
\usepackage{graphicx}
\usepackage{bm}% bold math
\usepackage{amssymb}
\usepackage{setstack}
\begin{document}
%-----------------------------------------------
\title[Analogue quantum gravity phenomenology from a two-component BEC]
{
Analogue quantum gravity phenomenology from a two-component Bose--Einstein condensate\\
}%
%-----------------------------------------------

\author{%
Stefano Liberati \footnote[1]{liberati@sissa.it}
}
\address{International School for Advanced
  Studies, Via Beirut 2-4, 34014 Trieste,
  Italy, \\ and INFN, Trieste}

\author{
Matt Visser \footnote[2]{matt.visser@mcs.vuw.ac.nz}
and
Silke Weinfurtner \footnote[3]{silke.weinfurtner@mcs.vuw.ac.nz}
}
\address{School of Mathematics, Statistics, and Computer Science, \\
Victoria University of Wellington, \\
P.O. Box 600, Wellington, New Zealand}

%-----------------------------------------------
\begin{abstract}
%-----------------------------------------------

We present an analogue emergent spacetime that reproduces the 
salient features of the most common ans\"atze used for quantum
gravity phenomenology.  We do this by investigating a system
of two coupled Bose--Einstein condensates. This system can be tuned to
have two ``phonon'' modes (one massive, one massless) which
share the same limiting speed in the hydrodynamic approximation [Phys.~Rev.~{\bf D72} (2005) 044020,  gr-qc/0506029; cond-mat/0409639]. The system nevertheless possesses (possibly non-universal) Lorentz violating terms once ``quantum pressure'' becomes important. 
We investigate the physical interpretation of the relevant fine-tuning conditions, and discuss the possible lessons and hints that this analogue spacetime could provide for the phenomenology of real physical quantum gravity. In particular we show that the effective field theory of quasi-particles in such an emergent spacetime does not exhibit the so called ``naturalness problem". 

%--------------------------------------------------------------

\vspace*{2.5mm}%
\vspace*{2.5mm}% 

%\bigskip

\noindent
{gr-qc/0510125; 30 October 2005; 26 February 2006; \LaTeX-ed \today.}
\\
{Final version --- to appear in Classical and Quantum Gravity}

%-----------------------------------------------
\end{abstract}
%-----------------------------------------------

\vspace{ -3mm}

%\pacs{04.20.-q; 04.20.Cv; ??????}

%-----------------------------------------------
\maketitle
%-----------------------------------------------
\def\d{{\mathrm{d}}}
%-------------------------------------------------------------------------
\def\implies{\Rightarrow}
%--------------------------------------------------------------------------
%------------------------------------------
%------------------------------------------
%------------------------------------------
% Lots of peculiar definitions
%------------------------------------------
\newcommand{\norm}[1]{\left\Vert#1\right\Vert}
\newcommand{\betrag}[1]{\left\vert#1\right\vert}
\newcommand{\bra}[1]{\langle #1\vert}
\newcommand{\ket}[1]{\arrowvert #1 \rangle}
\newcommand{\braket}[1]{\langle #1\rangle}
\newcommand{\kin}{-\frac{\hbar^2}{2m} \, \nabla^2 \,}
\newcommand{\oppdag}[2]{\hat{#1}^\dagger(t,\vec{#2})}
\newcommand{\opp}[2]{\hat{#1}(t,\vec{#2})}
\newcommand{\bog}{\hat{\Psi}(t,\vec{x})=\Phi(t,\vec{x})
  \, + \, \varepsilon \hat{\psi}(t,\vec{x})\, + \,
  \ldots} \newcommand{\zeit}{i \,\hbar \,
  \frac{\partial}{\partial t} \,}
\newcommand{\mad}{\sqrt{\rho(t,\vec{x})}e^{i\theta(t,\vec{x})}e^{-\frac{i\mu
      t}{\hbar}}}
%------------------------------------------
\def\half{{1\over2}} 
\def\L{{\mathcal L}}
\def\S{{\mathcal S}} 
\def\d{{\mathrm{d}}} 
\def\x{{\mathbf x}} 
\def\v{{\mathbf v}} 
\def\im{{\rm i}}
\def\etal{{\emph{et al\/}}} 
\def\det{{\mathrm{det}}}
\def\tr{{\mathrm{tr}}} 
\def\ie{{\emph{i.e.}}}
\def\bnabla{\mbox{\boldmath$\nabla$}}
\def\Box{\kern0.5pt{\lower0.1pt\vbox{\hrule height.5pt
      width 6.8pt \hbox{\vrule width.5pt height6pt
        \kern6pt \vrule width.3pt} \hrule height.3pt
      width 6.8pt} }\kern1.5pt}
\def\HRULE{{\bigskip\hrule\bigskip}}
%------------------------------------------
\def\Schrodinger{Schr\"odinger}
%------------------------------------------

%--------------------------------------------------------------
\def\d{{\mathrm{d}}} 
\def\g{{\mbox{\sl g}}}
\def\Box{\nabla^2}
\def\d{{\mathrm d}}
 \def\R{{\rm I\!R}}%
%--------------------------------------------------
\def\ie{{\em i.e.\/}}
 \def\eg{{\em e.g.\/}}
\def\etc{{\em etc.\/}}
 \def\etal{{\em et al.\/}}
%--------------------------------------------------------------
%--------------------------------------------------------------
%--------------------------------------------------------------
\section{Introduction}%
\label{sec:qg-ph}%
%--------------------------------------------------------------

The search for a quantum theory encompassing gravity has been a major issue in theoretical
physics for the last 60 years. Nonetheless, until recently quantum gravity was largely relegated to the
realm of speculation due to the complete lack of observational or
experimental tests. In fact the traditional scale of
quantum gravitational effects, the Planck scale $M_{\rm
  Pl}=1.2\times10^{19}$ GeV/$c^2$, is completely out of reach for any
experiment or observation currently at hand. This state of affairs had
led the scientific community to adopt the 
``folklore'' that testing quantum
gravity was completely impossible.
However the last decade has seen a dramatic change in
this respect, and nowadays one can encounter a growing
literature dealing with tests of possible predictions of various
quantum gravity models~\cite{LIV, Jacobson:2002hd, TeV-QG, others}. This
field goes generically under the name of ``quantum
gravity phenomenology".

Among the several generic predictions associated with quantum
gravity models, the possibility that Planck-scale physics might induce
violations of Lorentz invariance has played a particularly important role~\cite{LIV, Jacobson:2002hd}. 
Generically any possible ``discreteness'' or ``granularity'' of spacetime at the Planck scale seems
incompatible with strict Lorentz invariance (although some
particular quantum gravity theories might still preserve it, see
\emph{e.g.}~\cite{Rovelli}) as larger and larger boosts expose shorter and shorter distances. 
Actually we now have a wealth of theoretical studies --- for example in the context of string field theory~\cite{KS89, Damour:1994zq}, spacetime foam scenarios~\cite{GAC-Nat},
semiclassical calculations in loop quantum gravity~\cite{GP,loopqg}, DSR models~\cite{DSR, DSRint, Judes}, or
non-commutative
geometry~\cite{Hayakawa,Mocioiu:2000ip,Carroll:2001ws,Anisimov:2001zc}, just to cite a few --- all leading to high energy violations of Lorentz invariance.

Interestingly most investigations, even if they arise from quite different 
fundamental physics, seem to converge on the prediction that the breakdown of Lorentz invariance can generically become manifest in the form of modified dispersion relations exhibiting extra energy-dependent  or momentum-dependent terms, apart from the usual quadratic one occurring  in the Lorentz
invariant dispersion relation:
\begin{equation}
E^2=m^2\;c^4+p^2\;c^2.
\end{equation}
In the absence of a definitive theory of quantum gravity it became
common to adopt,  in most of the literature seeking to put these predictions to observational test, a purely
phenomenological approach, \ie,~one that modifies the dispersion relation by adding some generic momentum-dependent (or
energy-dependent) function $F(p,c,M_{\rm Pl})$ to be expanded in powers of the dimensionless quantity $p/(M_{\rm Pl}\; c)$. Hence the ansatz reads:
\begin{eqnarray}
E^2&=&m^2\;c^4+p^2 \;c^2 +F(p,c,M_{\rm Pl});
\\
&=& m^2\;c^4+p^2 \;c^2 + \sum_{n=1}^\infty \varpi_n \; p^n;
\\
&=&
m^2\;c^4+p^2\;c^2+c^4\left\{\eta_1\, M_{\rm Pl}\, p/c+\eta_2\,p^2/c^2+\sum_{n\geq3}
\eta_n\,\frac{(p/c)^{n}}{M^{n-2}_{\rm Pl}}\right\};
\label{eq:mod-disp}
\end{eqnarray}
where all the $\varpi_n$ carry appropriate dimensions, and in contrast the $\eta_n$ are chosen to be dimensionless. Since these dispersion relations are not Lorentz
invariant, it is necessary to specify the particular inertial frame in which they are given, and generally one chooses the CMB frame. Finally note that we have assumed rotational invariance and hence Lorentz violation only in the boost subgroup. This is motivated by the idea that Lorentz violation may arise in quantum gravity from the presence of a short distance cutoff. Moreover it is very difficult to conceive a framework where a breakdown of rotational invariance does not correspond to a violation of  boost invariance as well.

Of course merely specifying a set of dispersion relations is not always enough to place significant constraints --- as most observations need at least some assumption on the dynamics for their interpretation. In fact most of the available constraints are extracted from some assumed model ``test theory". 
Although several alternative scenarios have been considered in the literature, 
so far the most commonly explored avenue is an effective field theory (EFT) approach (see \eg,~\cite{jlm-ann} for a review focussed on this framework, and ~\cite{jlm-notes, jlm-qgp, jlm-limits, jlm-comments, jlm-nature,  jlm-tev, jlm-bloomington} for some of the primary literature). The main reasons for this choice can be summarized in the fact that we are very familiar with this class of theories, and that it is a widely accepted idea (although not unanimously accepted, see \eg~\cite{GAC-crit}) that \emph{any} quantum gravity scenario should admit a suitable EFT description at low energies.~\footnote{It is true that, \eg, non-commutative
geometry can lead to EFTs with problematic IR/UV mixing,
however this more likely indicates a physically unacceptable feature
of such specific models, rather than a physical limitation of EFT.} All in all, the standard model of particle physics and general relativity itself, (which are presumably not fundamental theories) are EFTs, as are most models of condensed matter systems at appropriate length and energy scales. Even ``fundamental'' quantum gravity candidates such as string theory admit an EFT description at low energies (as perhaps most impressively verified in the calculations of black hole entropy and
Hawking radiation rates). 

The EFT approach to the study of Lorentz violations has been remarkably useful in the last decade. Nowadays the best studied theories incorporating Lorentz violations are EFTs where Lorentz violations  are associated  either with renormalizable Lorentz-violating operators (mass dimension four or less), or sometimes with higher-order Lorentz-violating operators (mass dimension five and six or greater, corresponding to order $p^3$ and $p^4$ and higher deviations in the dispersion relation~(\ref{eq:mod-disp})). The first approach is generally known as the Standard Model Extension~\cite{ESM}, while the second has been formalized by Myers and Pospelov~\cite{MP} in the form of QED with dimension five Lorentz-violating operators (order $p^3$ deviations in the dispersion relation of equation~(\ref{eq:mod-disp})). In both cases extremely accurate constraints have been obtained using a combination of  experiments as well as observation (mainly in high energy astrophysics). See \eg,~\cite{LIV, jlm-ann}.

In the present article we wish to focus on the non-renormalizable EFT with Lorentz violations developed in~\cite{MP}, and subsequently studied by several authors.  In spite of the remarkable success of this proposal as a ``test theory'', it is interesting to note that there are still significant open issues concerning its theoretical foundations. In particular, let us now focus on  two aspects of this approach that have spurred some debate among the quantum gravity phenomenology
community.

\paragraph{The naturalness problem:}
%------------------------------------------

Looking at the dispersion relation (\ref{eq:mod-disp}) it might seem that the deviations linear and quadratic in $p$ are not Planck suppressed, and hence are always dominant (and grossly incompatible with observations).
However one might hope that there will be some other characteristic QFT mass scale $\mu\ll M_{\rm Pl}$ (\ie, some particle physics mass scale) associated with the Lorentz symmetry breaking which might enter in the lowest order dimensionless coefficients
$\eta_{1,2}$, which will be then generically suppressed by appropriate ratios of this characteristic mass to the Planck mass. 
Following the observational leads one might then assume behaviours like $\eta_1\propto (\mu /M_{\rm Pl})^{\sigma+1}$, and $\eta_2\propto (\mu/M_{\rm Pl})^\sigma$ where $\sigma\geq 1$ is some positive power (often taken as one or two). Meanwhile no extra Planck suppression is assumed for the higher order $\eta_n$ coefficients which are then naturally of order one. Note that such an ansatz assures that the Lorentz violation term linear in the particle momentum in equation~(\ref{eq:mod-disp}) is always subdominant with respect to the quadratic one, and that the Lorentz violating term cubic in the momentum is the less suppressed of the higher order ones.~\footnote{Of course this is only true provided there is no symmetry like parity that automatically cancels all the terms in odd powers of the momentum. In that case the least suppressed Lorentz-violating term would be that quartic in the momentum.}  If this is the case one will have two distinct regimes: For low momenta $p/(M_{\rm Pl}c) \ll (\mu/M_{\rm Pl})^\sigma$ the lower-order (quadratic in the momentum) deviations in~(\ref{eq:mod-disp}) will dominate over the higher-order (cubic and higher) ones, while at high energies $p/(M_{\rm Pl}c) \gg (\mu/M_{\rm Pl})^\sigma$ the higher order terms (cubic and above in the momentum) will be dominant.

The naturalness problem arises because such a line of reasoning does not seem to be well justified within an EFT framework. In fact we implicitly assumed that there are no extra Planck suppressions hidden in the dimensionless coefficients $\eta_n$ with $n\geq 3$. Indeed we cannot justify why \emph{only} the dimensionless coefficients of the $n\leq 2$ terms should be suppressed by powers of the small ratio $\mu/M_{\rm Pl}$.  Even worse, renormalization group arguments seem to imply that a similar mass ratio, $\mu/M_{\rm Pl}$ would implicitly be present in \emph{all} the dimensionless $n\geq3$ coefficients --- hence suppressing them even further, to the point of complete undetectability.  Furthermore it is easy to show~\cite{Collins} that, without some protecting symmetry,~\footnote{A symmetry which could play a protective role for the lowest-order operators as indeed been suggested. In~\cite{Pospelov-Nibbelink} it was shown that the dual requirements of supersymmetry and gauge invariance permit one to add to the SUSY standard model only those operators corresponding to $n\geq3$ terms in the dispersion relation. However it should be noted that in~\cite{Pospelov-Nibbelink} the $\eta_3$ coefficient carries a further suppression of order $m^2/M^2$ when compared to~(\ref{eq:mod-disp}).} it is generic that radiative corrections due to particle interactions in an  EFT with only Lorentz violations of order $n\geq 3$ in (\ref{eq:mod-disp}) for the free particles, will generate $n=1$ and $n=2$ Lorentz violating terms in the dispersion relation which will then be dominant.  

\paragraph{The universality issue:}
%---------------------------------------
The second point is not so much a problem,
as an issue of debate as to the best strategy to adopt. In dealing with situations with multiple particles one has to choose between the case of universal
(particle-independent) Lorentz violating coefficients
$\eta_n$, or instead go for a more general ansatz and
allow for particle-dependent coefficients; hence
allowing different magnitudes of Lorentz symmetry
violation for different particles even when considering
the same order terms (same $n$) in the momentum
expansion. The two choices are equally represented in the
extant literature (see \emph{e.g.}~\cite{GAC-Pir} and \cite{Jacobson:2002hd} for the two alternative ans\"atze), but it would be interesting to
understand how generic this universality might be, and what sort of processes might induce non-universal  Lorentz violation for different particles.

\vskip 20 pt

To shed some light on these issues it would definitely be
useful to have something that can play the role of test-bed for some of the ideas related to the emergence and form of the modified dispersion relations of equation~\eref{eq:mod-disp}.  In
this regard,  herein we will consider an analogue model of
emergent spacetime, that is, a condensed matter system which admits
excitations whose propagation mimics that of quantum
fields on a curved spacetime~\cite{LivRev}.\footnote{For other rather distinct views on ``emergent spacetime'', with rather different aims, see for instance the articles by C D Froggatt and H B Nielsen~\cite{Nielsen}, by J D Bjorken~\cite{Bjorken},  and by R B Laughlin~\cite{Laughlin}.}  Indeed it is well known
that the discreteness at small scales of such systems
shows up exactly via modified dispersion relations of the
kind described by equation \eref{eq:mod-disp}, and one may hope that the complete
control over the microscopic (trans-Planckian) physics in
these systems would help understanding the nature of the
issues discussed above~\cite{BLV, Breakdown}.
For example, we remind the reader that the Bogolubov quasi-particle spectrum for excitations of a Bose--Einstein condensate
\begin{equation}
\omega^2 = c_s^2 \; k^2 +   c_s^2 \; k^4/K^2,
\label{bogo-disp}
\end{equation}
where $c_s$ is the speed of sound
and $K$ is determined by the effective Compton wavelength~\cite{BLV,Breakdown} 
\begin{equation}
K=\frac{2\pi}{\lambda_\mathrm{Compton}}=2\pi \frac{mc_s}{\hbar},
\end{equation}
\ie, it is set by the the mass of the fundamental bosons forming the BEC. Note also that we do not encounter odd powers of the momentum in (\ref{bogo-disp}), as the system is by construction invariant under parity. 

This system (which has been extensively studied as an analogue model of gravity~\cite{BLV, Breakdown, Garay}, in particular with reference to the simulation of black holes via supersonic flows), provides a simple and explicit example of the high-energy breakdown of ``Lorentz invariance"~\cite{BLV,Breakdown} with a dispersion relation of the form~(\ref{eq:mod-disp}) which interpolates between a low-energy ``massless'' relativistic regime
\begin{equation}
\omega^2 \approx c_s^2 k^2, \qquad  (k \ll K);
\end{equation}
and a non-relativistic, approximately Newtonian, high-energy regime
\begin{equation}
\omega \approx c_s k^2/ K, \qquad  (k \gg K).
\end{equation}
Here $K$ is hence the scale of the violation of the Lorentz invariance and as such  in the language of quantum gravity phenomenology it is the analogue of the Planck scale.~\footnote{Let us stress that while it is a standard assumption in quantum gravity phenomenology to identify the scale of Lorentz violation with the Planck scale, it is not a priori  necessary that the two must exactly coincide. In the discussion below, what we shall call the effective or analogue ``Planck scale'' has always to be interpreted as the scale of Lorentz breaking;  given that this is the only relevant scale in our discussion.}

Unfortunately it is easy to realize that this particular system is too simple in order to mimic the salient issues of quantum gravity phenomenology. Actually all of the analogue models currently at hand have a problem in this sense, either because they
do not provide the same dispersion relations for
different excitations even at low frequencies (in the words of~\cite{VW1, VW2}, ``no
mono-metricity''), or because they deal with just one single kind of
excitation (like the 1-BEC model just discussed)  --- in which case it is impossible to say anything
about either naturalness or universality. 
[Regarding the naturalness problem, if there is only one type of excitation 
present in the system then one
  cannot identify a $\eta_2$ modification at the quadratic ($p^2$)
  order, given that it shows up only via differences in
  the limit speed when comparing different particles.]
  
However it was recently realized that experimentally available systems of two coupled BECs are much richer in their spectrum, and allow the simulation of an analogue spacetime where two different particles coexist and interact through mode-mixing. In particular in~\cite{VW1,VW2} it was shown that for a 2-BEC system there are ways of modifying the excitation spectrum (\ref{bogo-disp}) in order to add a ``mass" term, and the analysis of the present article will build on those two articles. (For a somewhat related though distinct application of 2-BEC systems to analogue models, see~\cite{Fischer}.)

We shall consider the special case of a homogeneous
two-component BEC subject to laser-induced coupling. This system 
exhibits a rich spectrum of excitations, which can
be viewed as two interacting phonon modes (two quasiparticle modes). We study
the conditions required for these two phonon modes to
share the same ``special relativity'' metric in the hydrodynamic
limit (effectively the low-energy limit), and find that (in this limit) the
two phonons respectively exhibit   a massive and
massless Lorentz invariant dispersion relation. 
We then consider the high energy limit of the system, that is, situations where the  
 ``quantum pressure'' term, which in a 1-BEC system is is at the origin of the $k^4$ Lorentz violation in (\ref{bogo-disp}), can no longer be neglected. 

Though much of the underlying physics is similar to that of~\cite{VW1, VW2}, the central thrust of the argument is different. In those articles one was always working in the hydrodynamic limit, often with inhomogeneous backgrounds, and seeking to extract a curved spacetime metric. In the current article we are working with homogeneous backgrounds, staying as close as possible to ``special relativity'', and specifically probing the possible breakdown of Lorentz invariance by going beyond the hydrodynamic approximation. Thus many of the issues that normally are central to the discussion of analogue models (such as the existence of a curved effective spacetime, analogue horizons, causal structure, analogue Hawking radiation, the simulation of cosmological spacetimes, and the like~\cite{abh, silke, Schutzhold:cosmic}), here are of at best peripheral interest. The analogue spacetimes we are interested in are all flat.
Finally we also stress that our main thrust is here to (eventually) learn lessons about how real physical quantum gravity might work --- we are not particularly concerned about the experimental laboratory  realizability of our specific analogue system. (Readers more interested in specific condensed-matter aspects of 2-BEC systems  and their excitations might consult, for instance, references~\cite{trippenbach, jenkins,Toms},  and references therein.)

The rest of the paper will be organized as follows. In section \ref{sec:bec} we shall describe the general equations for the propagation of excitations of the 2-BEC background, while section \ref{sec:heal} then discusses the notion of ``healing length'' in a 2-component BEC. Then in section \ref{sec:hydro} we shall explicitly consider the low energy limit when the quantum potential (\ie, the ultraviolet physics due to the atomic nature of the condensate) is negligible. This limit will allow us to identify for which combination of the microscopic parameters this system is indeed an analogue model of gravity, \ie, it is characterized by a single metric for all the excitations. In section \ref{sec:qp} we then move on to explore how UV effects, embodied by the quantum potential, introduce Lorentz violations, and in section \ref{sec:qgp} we study the implications for quantum gravity phenomenology. In the final discussion, section \ref{sec:discuss}, we consider the lessons one can draw from this analogue model about how low energies Lorentz violations can be protected in realistic situations.

%--------------------------------------------------------------
\section{Phonons in two-component BECs}%
\label{sec:bec}%
%--------------------------------------------------------------
Two BECs interacting with each other, and coupled by a laser-driven coupling, can usefully be described by the pair of Gross--Pitaevskii equations~\cite{VW1, VW2}:
\begin{eqnarray}
\fl
 i \, \hbar \, \partial_{t} \Psi_{A} &=& \left[
   -\frac{\hbar^2}{2\,m_{A}} \nabla^2 + V_{A}-\mu + U_{AA}
   \, \betrag{\Psi_{A}}^2 + U_{AB} \betrag{\Psi_{B}}^2
   \right] \Psi_{A} + \lambda \, \Psi_{B} \, , \label{2GPEa}\\ 
\fl
i \,
 \hbar \, \partial_{t} \Psi_{B} &=& \left[
   -\frac{\hbar^2}{2\,m_{B}} \nabla^2 + V_{B}-\mu + U_{BB}
   \, \betrag{\Psi_{B}}^2 + U_{AB} \betrag{\Psi_{A}}^2
   \right] \Psi_{B} +\,\lambda \, \Psi_{A} \,  \label{2GPEb}.
\end{eqnarray}
We permit $m_A\neq m_B$ in the interests of generality, (although $m_A\approx m_B$ in all currently realizable experimental systems), and note that $\lambda$ can take either sign without restriction. In the specific idealized case $\lambda=0$ and $U_{AB}=0$ we have two inter-penetrating but non-coupled BECs, each of which separately exhibits a Bogolubov spectrum with distinct values of the Lorentz breaking scale $K$ (since $m_A\neq m_B$). Once the BECs interact the spectrum becomes more complicated, but that is exactly the case we are interested in for this article.

To analyze the excitation spectrum we linearize around some background using:
\begin{equation}
\Psi_{X}= \sqrt{\rho_{X0}+ \varepsilon \, \rho_{X1} }\,
e^{i(\theta_{X0}+ \varepsilon \, \theta_{X1} )}
\quad\hbox{for}\quad X=A,B \, .
\end{equation}
Because we are primarily interested in looking at deviations from special relativity (SR), we take our background to be homogeneous [position-independent], time-independent, and at rest [$\vec v_{A0}=0=\vec v_{B0}$]. That is, we will be dealing with an analogue of flat Minkowski spaccetime. We also set the background phases equal to each other, $\theta_{A0}=\theta_{B0}$. This greatly simplifies the technical computations.\footnote{For some of the additional complications when background phases are unequal see~\cite{VW1}.}
Then the linearized Gross--Pitaevskii equations imply:
\begin{equation} \label{soundwave1}
\dot{\theta}_{A1}= 
- \frac{\tilde{U}_{AA}}{\hbar} \, \rho_{A1}
- \frac{\tilde{U}_{AB}}{\hbar} \,\rho_{B1}
+\frac{\hbar}{2m_A}\hat Q_{A1}(\rho_{A1}),
\\ 
\end{equation}
\begin{equation}
\dot{\theta}_{B1}= 
-\frac{\tilde{U}_{BB}}{\hbar} \, \rho_{B1} 
-\frac{\tilde{U}_{AB}}{\hbar} \,\rho_{A1}
+\frac{\hbar}{2m_B}\hat Q_{B1}(\rho_{B1}), \\ 
\end{equation}
and
\begin{equation} \label{soundwave2}
\dot{\rho}_{A1}= 
- \frac{\hbar}{m_{A}}
\,\rho_{A0} \nabla^2 \theta_{A1} +
\frac{2\lambda}{\hbar}\sqrt{\rho_{A0} \, \rho_{B0}} \;
(\theta_{B1}-\theta_{A1}), 
\end{equation}
\begin{equation}
\dot{\rho}_{B1}= 
- \frac{\hbar}{m_{B}} \, \rho_{B0} \nabla^2
\theta_{B1} +
\frac{2\lambda}{\hbar}\sqrt{\rho_{A0}\,\rho_{B0}} \;
(\theta_{A1}-\theta_{B1}) .
\end{equation}
Here we have defined
\begin{eqnarray}
\tilde{U}_{AA} &= U_{AA} -\frac{\lambda}{2}
\frac{\sqrt{\rho_{B0}}}{\sqrt{\rho_{A0}}^{3}}
&= 
U_{AA}-{\lambda\sqrt{\rho_{A0}\rho_{B0}}\over2} {1\over \rho_{A0}^2} , 
\nonumber
\\
\tilde{U}_{BB} &= U_{BB} -\frac{\lambda}{2}
\frac{\sqrt{\rho_{A0}}}{\sqrt{\rho_{B0}}^{3}} 
&= 
U_{BB}-{\lambda\sqrt{\rho_{A0}\rho_{B0}}\over2} {1\over \rho_{B0}^2}, 
\\
\tilde{U}_{AB} &= U_{AB}+ \frac{\lambda}{2}  \;
\frac{1}{\sqrt{\rho_{A0} \, \rho_{B0} }}
&= 
U_{AB}+{\lambda\sqrt{\rho_{A0}\rho_{B0}}\over2} {1\over \rho_{A0}\rho_{B0}},
\nonumber
\end{eqnarray}
and furthermore we define 
$\hat Q_{X1}$ as the second-order
differential operator obtained from linearizing the
quantum potential:
\begin{eqnarray}
 V_{\rm Q}(\rho_X) 
&\equiv&  
-  {\hbar^2\over2m_X} \left( {\nabla^2\sqrt{\rho_X}\over\sqrt{\rho_X}}\right) 
=
-  {\hbar^2\over2m_X} \left( {\nabla^2\sqrt{\rho_{X0}+\varepsilon\rho_{X1}}
\over\sqrt{\rho_{X0}+\varepsilon\rho_{X_1}} }\right) 
\\
&=& 
-  {\hbar^2\over2m_X} 
\left(  \hat Q_{X0}(\rho_{X0}) + \varepsilon \;\hat Q_{X1} (\rho_{X1}) \right).
\end{eqnarray}
The quantity $ \hat Q_{X0}(\rho_{X0})$ corresponds to the background value of the quantum pressure, and contributes only to the background equations of motion --- it does not affect the fluctuations. Now in a general background
\begin{equation}
\hat Q_{X1} (\rho_{X1})= {1\over2} \left\{ 
{ (\nabla\rho_{X0})^2-(\nabla^2\rho_{X0})\rho_{X0}\over\rho_{X0}^3} 
- {\nabla \rho_{X0}\over\rho_{X0}^2} \nabla 
+ {1\over\rho_{X0}} \nabla^2
\right\} \rho_{X1}.
\end{equation}
Given the homogeneity of the background appropriate for the current discussion this simplifies to
\begin{equation}
\hat Q_{X1} (\rho_{X1})=  {1\over2\rho_{X0}} \nabla^2 \rho_{X1}.
\end{equation}

The set of first-order partial differential equations relating the phase fluctuations and density fluctuations can be written in a
more concise matrix form. First let us define
 a set of $2\times2$ matrices, starting
with the coupling matrix 
\begin{equation}
\hat{\Xi}=\Xi+\hat X, 
\end{equation}
where
\begin{equation}
\Xi=\frac{1}{\hbar} \left[
\begin{array}{rr}
\tilde{U}_{AA} &  \tilde{U}_{AB}  \\ \tilde{U}_{AB} &
\tilde{U}_{BB}  \\
\end{array}
\right],
\end{equation} 
and
\begin{eqnarray}
%\fl
\hat X &=&
 -{\hbar\over2} \left[\begin{array}{cc} {\hat Q_{A1}\over m_A} & 0
    \\ 0 & {\hat Q_{B1}\over m_B}\end{array}\right]
% \\
%&=&
=
-{\hbar\over4} \left[\begin{array}{cc} {1\over m_A\;\rho_{A0}} & 0
    \\ 0 & {1\over m_B\;\rho_{B0}} \end{array}\right]\nabla^2
=
 - X \; \nabla^2
\, .
\end{eqnarray}
A second coupling matrix can be introduced as
\begin{equation}
\Lambda= -\frac{2\lambda\;\sqrt{\rho_{A0}\,\rho_{B0}} }{\hbar}
\left[
\begin{array}{rr}
+1 & -1 \\ -1 & + 1 \\
\end{array}
\right] \, .
\end{equation}
It is also useful to introduce the mass-density matrix $D$
\begin{equation}
D=\hbar \left[
\begin{array}{cc}
{\rho_{A0}}/{m_{A}}  & 0 \\ 0 &
{\rho_{B0}}/{m_{B}}   \\
\end{array}
\right] .
\end{equation}
Now define two column vectors  ${\bar{\theta}} = [\theta_{A1},\theta_{B1}]^T$
and  ${\bar{\rho}} = [\rho_{A1},\rho_{B1}]^T$.\\

Collecting terms into a $2\times2$ matrix equation, the
equations for the phases (\ref{soundwave1}) and densities
(\ref{soundwave2}) become
\begin{equation} \label{thetavecdot}
\dot{\bar{\theta}}=  -\,\hat\Xi  \; \bar{\rho},
\end{equation} 
\begin{equation} \label{rhovecdot}
\dot{\bar{\rho}}= \, - D  \; \nabla^2
\bar{\theta}  + \Lambda 
\bar{\theta} \, .
\end{equation} 
Equation (\ref{thetavecdot}) can now be used to eliminate $\dot{\bar{\rho}}$ in equation
(\ref{rhovecdot}), leaving us with a single matrix
equation for the perturbed phases:
\begin{equation} \label{phaseequation2}
 \partial_{t} \left(\hat\Xi^{-1} \; \dot{\bar{\theta}} \, \right) = 
  D \; \nabla^2 \bar{\theta} - \Lambda
  \; \bar{\theta}
\end{equation}
This is an integro-differential equation [since $\hat\Xi$ is a matrix of differential operators] which is a
second-order differential equation in time, but an
integral equation (equivalently, an infinite-order
differential equation) in space.  

We now formally construct the operators $\hat \Xi^{1/2}$ and 
 $\hat \Xi^{-1/2}$ and use them to define a new set of variables
\begin{equation}
\tilde\theta = \hat \Xi^{-1/2} \;\bar\theta,
\end{equation}
in terms of which the wave equation becomes
\begin{equation}
\partial_t^2\tilde\theta = 
\left\{ \hat \Xi^{1/2}\; [ D \nabla^2 - \Lambda] \; \hat \Xi^{1/2} \right\} \tilde\theta,
\end{equation}
or more explicitly
\begin{equation}
\label{eq:last}
\partial_t^2\tilde\theta = 
\left\{ [\Xi- X \nabla^2]^{1/2}\;  [ D \nabla^2 - \Lambda] \; 
 [\Xi-X\nabla^2] ^{1/2} \right\} \tilde\theta.
\end{equation}
This is now a (relatively) simple PDE to analyze.  Note that whereas the objects $\hat \Xi^{1/2}$ and 
 $\hat \Xi^{-1/2}$  are $2\times2$ matrices whose elements are pseudo-differential operators,  as a practical matter we never have to descend to this level of technicality.
Indeed, it is computationally efficient to directly go to the 
eikonal limit where
\begin{equation}
\hat \Xi \to \Xi + X\; k^2.
\end{equation}
This finally leads to a dispersion relation of the form
\begin{eqnarray}
\label{fresnel}
&&\det\big\{ \omega^2 \; \mathbf{I} -
 [\Xi+ X k^2]^{1/2}\;  [ D k^2  + \Lambda] \; 
 [\Xi+ X k^2] ^{1/2}  \big\} =0\,,
\end{eqnarray}
and ``all'' we need to do for the purposes of this article, is to understand this quasiparticle excitation spectrum in detail.

%-----------------------------------------
\section{Healing length}
\label{sec:heal}
%------------------------------------------
Note that the differential operator $\hat Q_{X1}$ that underlies the origin of the
$X\, k^2$ contribution above is obtained by
linearizing the quantum potential
\begin{equation}
 V_{\rm Q}(\rho_X) \equiv  -  {\hbar^2\over2 m_X}
 \left( {\nabla^2 \sqrt{\rho_X}\over\sqrt{\rho_X}}\right)
\end{equation}
which appears in the Hamilton--Jacobi equation of the BEC
flow.  This quantum potential term is suppressed by the smallness of $\hbar$, the comparative largeness of $m_X$,  and
for sufficiently uniform density profiles. 
But of course in any real system the density of a BEC must go to zero at the boundaries of its
EM trap (given that $\rho_X=|\psi_X(\vec x,t)|^2$). 
In a 1-component BEC the healing length characterizes the minimal distance over which the order parameter goes from zero to its bulk value. If the
condensate density grows from zero to $\rho_0$ within a distance $\xi$ the
quantum potential term (non local) and the interaction energy (local)
are respectively $E_{\rm kinetic}\sim \hbar^2/(2m\xi^2)$ and $E_{\rm
  interaction}\sim 4\pi\hbar^2 a \rho_0/m$. These two terms are comparable when
 \begin{equation}
\xi=(8\pi \rho_0 a)^{-1/2},
  \label{heal}
 \end{equation}
where $a$ is the $s$-wave scattering length defined as
\begin{equation}
a = {m \; U_0\over4\pi \hbar^2}.
\end{equation}
Note that what we call $U_0$ in the above expression is just the coefficient of the non-linear self-coupling term in the Gross--Pitaevskii equation, \ie, just $U_{AA}$ or $U_{BB}$ if we completely decouple the 2 BECs ($U_{AB}=\lambda=0$). 

From the definition of the healing length it is hence clear that only for excitations with
wavelengths much larger than the healing length is the
effect of the quantum potential negligible. This is
called the hydrodynamic limit because the single--BEC
dynamics is described by the continuity and
Hamilton--Jacobi equations of a super-fluid, and its
excitations behave like massless phononic modes. In the case
of excitations with wavelengths comparable with the
healing length this approximation is no longer appropriate
and deviations from phononic behaviour will arise.  

Such a simple discrimination between different regimes is lost once one considers a system formed by two coupled Bose--Einstein condensates. In fact in this case one is forced to introduce a generalization of the healing $\xi$ length in the form of a ``healing matrix". Let us elaborate on this point. If we apply the same reasoning used above for the definition of the ``healing length'' to the 2-component BEC system (\ref{2GPEa}, \ref{2GPEb}) we again find a functional form like that of equation~(\ref{heal}) however we now have the crucial difference that both the density and the scattering length are replaced by matrices. In particular we generalize the scattering length $a$ to the matrix $\mathcal{A}$:
 \begin{equation}
 \mathcal{A} = {1\over4\pi\hbar^2} 
 \left[\begin{array}{cc}\sqrt{m_A}&0\\0&\sqrt{m_B}\end{array}\right]
 \;
 \left[\begin{array}{cc}
 \tilde U_{AA}&\tilde U_{AB}\\ \tilde U_{AB} &\tilde U_{BB} 
 \end{array}\right]
 \;
 \left[\begin{array}{cc}\sqrt{m_A}&0\\0&\sqrt{m_B}\end{array}\right],
 \end{equation}
Furthermore, from (\ref{heal}) a healing length matrix $Y$ can be defined by 
 \begin{equation}
 \fl
Y^{-2} = {2\over\hbar^2} \left[\begin{array}{cc}
\sqrt{\rho_{A0} m_A}&0\\0&\sqrt{\rho_{B0} m_B}\end{array}\right]
 \;
 \left[\begin{array}{cc}
 \tilde U_{AA}&\tilde U_{AB}\\ \tilde U_{AB} &\tilde U_{BB} 
 \end{array}\right]
 \;
 \left[\begin{array}{cc}
 \sqrt{\rho_{A0} m_A}&0\\0&\sqrt{\rho_{B0} m_B}\end{array}\right].
 \end{equation}
 That is, in terms of the matrices we have so far defined:
 \begin{equation}
 Y^{-2} = {1\over2} \;X^{-1/2} \; \Xi \; X^{-1/2};
 \qquad
 Y^2 = 2\; X^{1/2} \; \Xi^{-1} \; X^{1/2}.
 \end{equation}
 We can now define ``effective'' scattering lengths and healing lengths for the 2-BEC condensate as
\begin{equation}
a_\mathrm{eff} = {1\over2}\;\tr[{\mathcal{A}}] =
 {m_A \tilde U_{AA} + m_B \tilde U_{BB}\over 8\pi\hbar^2},
\end{equation}
and
\begin{equation}
\xi_\mathrm{eff}^2 = {1\over2}\;\tr[Y^2] = \tr[X \Xi^{-1}] 
= {\hbar^2 [\tilde U_{BB}/(m_A \rho_{A0}) + \tilde U_{AA}/(m_B \rho_{B0} )]\over
4 (\tilde U_{AA} \tilde U_{BB} - \tilde U_{AB}^2 ) }.
\end{equation}
That is
\begin{equation}
\xi_\mathrm{eff}^2 = 
{\hbar^2 [m_A \rho_{A0} \tilde U_{AA}+ m_B \rho_{B0} \tilde U_{BB}]
\over
4 m_A m_B \rho_{A0} \rho_{B0} \;(\tilde U_{AA} \tilde U_{BB} - \tilde U_{AB}^2 ) }.
\end{equation}
Note that if the two components are decoupled and tuned to be equivalent to each other,  then these effective scattering and healing lengths reduce to the standard one-component results.
We shall soon see that sometimes it is convenient to deal with explicit formulae involving the ``low level'' fundamental quantities such as $m_A$, $m_B$, $\tilde U_{XY}$, \etc, while often it is more convenient to deal with ``high level''  quantities such as $\xi_\mathrm{eff}$.

%--------------------------------------------------------------
\section{Hydrodynamic approximation}%
\label{sec:hydro}%
%--------------------------------------------------------------

We are now interested in investigating the most general
conditions (in the hydrodynamic limit) under which this
two-BEC system can describe two phononic modes
propagating over the same metric structure. The
hydrodynamic limit is equivalent to formally setting
$\hat X\to0$ so that $\hat \Xi \to \Xi$.  (That is, one is formally setting the healing length matrix to zero: $Y \to 0$. More precisely, all components of the healing length matrix are assumed small compared to other length scales in the problem.) 

%------------------------------------
\subsection{Fresnel equation}
%------------------------------------

The PDE (\ref{phaseequation2}) now takes the simplified
form:
 \begin{equation}
 \partial_{t}^2 {\bar{\theta}}  = +  [\Xi \;D ] \cdot
 \nabla^2 \bar{\theta}  - [\Xi \; \Lambda] \cdot
 \bar{\theta} .
\end{equation}
Since this is second-order in both time and space
derivatives, we now have at least the possibility of
obtaining an exact ``Lorentz invariance''.  
We can now
define the transformed variables
\begin{equation}
\tilde\theta = \Xi^{-1/2} \;\bar\theta,
\end{equation}
and
the matrices
\begin{equation}
\Omega^2 = \Xi^{1/2}\;\Lambda\;\Xi^{1/2}; \qquad C_0^2 =   \Xi^{1/2}\; D\; \Xi^{1/2}; 
\end{equation}
so that 
\begin{equation}
 \partial_{t}^2 {\tilde{\theta}}  = +  C_0^2  \nabla^2
 \bar{\theta}  -\Omega^2  \bar{\theta}.
\end{equation}
Then in momentum space
\begin{equation}
\omega^2 {\tilde{\theta}}  = \left\{ C_0^2 \;
k^2+\Omega^2\right\} \; \tilde{\theta}  \equiv H(k^2) \; 
\tilde{\theta},
\end{equation}
leading to the Fresnel equation
\begin{equation}
\det\{ \omega^2 \;\mathbf{I} - H(k^2) \} =0.
\end{equation}
That is
\begin{equation}
\omega^4 - \omega^2 \; \hbox{tr}[H(k^2)] + \det[H(k^2)] =
0,
\end{equation}
whence
\begin{equation}
\omega^2 = { \hbox{tr}[H(k^2)] \pm \sqrt{
    \hbox{tr}[H(k^2)]^2 - 4\; \det[H(k^2)] }\over 2}.
\label{eq:disp-rel-hydro}
\end{equation}
Note that the matrices $\Omega^2$, $C_0^2$, and $H(k^2)$ have now carefully been arranged to be 
\emph{symmetric}. This greatly simplifies the subsequent matrix algebra. Also note that the matrix $H(k^2)$ is a function of $k^2$; this will forbid the appearance of odd powers of $k$ in the dispersion relation --- as should be expected due to the parity invariance of the system.
%

%----------------------------------------
\subsection{Masses}
%----------------------------------------
%
We read off the ``masses'' by looking at the special case of
space-independent oscillations for which
\begin{equation}
 \partial_{t}^2 {\bar{\theta}}  = -\Omega^2 \; \bar{\theta}
 ,
\end{equation}
allowing us to identify the ``mass'' (more precisely, the natural oscillation
frequency) as
\begin{equation}
\hbox{``masses''} 
\propto \hbox{eigenvalues of}\;(\Xi^{1/2}\;\Lambda\; \Xi^{1/2})
 =  \hbox{eigenvalues of}\;(\Xi\;\Lambda).
\end{equation}
Since $\Lambda$ is a singular $2\times2$ matrix this
automatically implies
\begin{equation}
\omega_I^2=0; \qquad \omega_{II}^2 = \hbox{tr}\,(\Xi\;\Lambda).
\label{eq:masses}
\end{equation}
So we see that one mode will be a massless phonon while the other will have a non zero mass. Explicitly, in terms of the elements of the underlying matrices
\begin{equation}
\omega_I^2=0; \qquad  \omega_{II}^2 =
-\frac{2\sqrt{\rho_{A0}\,\rho_{B0}} \;\lambda }{\hbar^2}
\{ \tilde U_{AA} + \tilde U_{BB} - 2 \tilde U_{AB} \}
\label{eq:masses2}
\end{equation}
so that (before any fine-tuning or decoupling)
\begin{equation}
\fl 
\omega_{II}^2 = -\frac{2\sqrt{\rho_{A0}\,\rho_{B0}} \;\lambda
}{\hbar^2} \left\{ U_{AA} +  U_{BB} - 2 U_{AB}
-{\lambda\over2\sqrt{\rho_{A0}\,\rho_{B0}}}  \left[
  \sqrt{\rho_{A0}\over \rho_{B0}} 
+\sqrt{\rho_{B0}\over
    \rho_{A0}}\right]^2 \right\}.
\label{eq:m-ndiag}
\end{equation}
It is easy to check that this quantity really does have the physical dimensions of a frequency.

%------------------------------------------------------------
\subsection{Conditions for mono-metricity}
\label{sec:mono-metr}
%------------------------------------------------------------
%
We now want our system to be a perfect analogue of
special relativity. That is:
\begin{itemize}
\item We want each mode to have a quadratic dispersion
  relation;
\item We want each dispersion relation to have the same
  asymptotic slope.
\end{itemize}
In order to find under which conditions these requirements
are satisfied we will adopt the following strategy:
Let us start by noticing that the dispersion relation
(\ref{eq:disp-rel-hydro}) is of the form
\begin{equation}
\omega^2 = [\hbox{quadratic}_1] \pm
\sqrt{[\hbox{quartic}]}.
\end{equation}
The first condition implies that the quartic must be a
perfect square
\begin{equation}
[\hbox{quartic}] = [\hbox{quadratic}_2]^2,
\end{equation}
but then the second condition implies that the slope of
this quadratic must be zero. That is
\begin{equation}
[\hbox{quadratic}_2](k^2) = [\hbox{quadratic}_2](0),
\end{equation}
and so
\begin{equation}
[\hbox{quartic}](k^2) =[ \hbox{quartic}](0) 
\end{equation}
must be constant independent of $k^2$, so that the
dispersion relation is of the form
\begin{equation}
\omega^2 = [\hbox{quadratic}_1](k^2) \pm[\hbox{quadratic}_2](0).
\end{equation}
Note that this has the required form (two hyperbolae with
the same asymptotes, and possibly different
intercepts). Now let us implement this directly in terms of the matrices
$C_0^2$ and $M^2$.

\paragraph{Step 1:}
Using the results of the appendix, specifically equation (\ref{E:A-2-matrices}):
\begin{eqnarray}
\det[H^2(k)] &=& \det[\Omega^2 + C_0^2\; k^2 ] \\ &=&
\det[\Omega^2] - \tr\left\{ \Omega^2\; \bar C_0^2\right\} \; k^2 + \det[C_0^2]\;(k^2)^2.
\end{eqnarray}
(This holds for any linear combination of $2\times2$
matrices. Note that we apply trace reversal to the squared matrix $C_0^2$,  we do not trace reverse and then square.) 
Since in particular $\det[\Omega^2]=0$,  we have:
\begin{equation}
\det[H^2(k)] = - \tr\left\{ \Omega^2\;\bar C_0^2 \right\} \; k^2 + \det[C_0^2]\;(k^2)^2.
\end{equation}

\paragraph{Step 2:}
Now consider the discriminant (the quartic)
\begin{eqnarray}
\fl \qquad \hbox{quartic} &\equiv&
\hbox{tr}[H(k^2)]^2 - 4\; \det[H(k^2)] 
\\ 
\fl &=& (\tr[\Omega^2]+\tr[C_0^2]\; k^2)^2 - 4 \left[
    -\tr\left\{ \Omega^2\;\bar C_0^2 \right\} \; k^2 + \det[C_0^2]\;(k^2)^2 \right]
\qquad
  \\ 
\fl &=& \tr[\Omega^2]^2 + \{2 \tr[\Omega^2]\tr[C_0^2]+4
  \tr\left\{ \Omega^2\;\bar C_0^2 \right\} \} k^2 
\nonumber \\ &&\qquad\qquad\qquad\qquad
  + \left\{ \tr[C_0^2]^2 - 4\det[C_0^2]\right\} (k^2)^2
  \\
\fl &=& \tr[\Omega^2]^2 +
 2 \{2 \tr\left\{ \Omega^2\; C_0^2 \right\} - \tr[\Omega^2]\tr[C_0^2] \} k^2 
\nonumber \\ &&\qquad\qquad\qquad\qquad
  + \left\{ \tr[C_0^2]^2 - 4\det[C_0^2]\right\} (k^2)^2.
\end{eqnarray}
So in the end the two conditions above for mono-metricity
take the form
\begin{equation}
\hbox{mono-metricity}\iff\left\{ 
\begin{array}{l}
\tr[C_0^2]^2 - 4\;\det[C_0^2] =0;
\\
2 \tr\left\{ \Omega^2\; C_0^2 \right\} -  \tr[\Omega^2] \;\tr[C_0^2]= 0.
\end{array}\right .
\end{equation}
Once these two conditions are satisfied the dispersion
relation is
\begin{equation}
\omega^2 = { \hbox{tr}[H(k^2)] \pm \tr[\Omega^2] \over 2} =
      {\tr[\Omega^2]\pm\tr[\Omega^2] + \tr[C_0^2]\; k^2\over 2}
\end{equation}
whence
\begin{equation}
\omega_1^2 = {1\over2} \tr[C_0^2]\; k^2=c_0^2k^2
\qquad
\omega_2^2 = \tr[\Omega^2] + {1\over2} \tr[C_0^2]\; k^2=\omega_{II}^2+c_0^2k^2,
\end{equation}
as required. One mode is massless, one massive with
exactly the ``mass'' previously deduced. One can now define the quantity
\begin{equation}
m_{II} = \hbar \omega_{II}/c_0^2,
\end{equation}
which really does have the physical dimensions of a mass.

%-----------------------------------------------------------
\subsection{Interpretation of the mono-metricity conditions}
%-----------------------------------------------------------
\label{S:C1C2}
%-----------------------------------------------------------

But now we have to analyse the two simplification
conditions
\begin{equation}
C1:\qquad \tr[C_0^2]^2 - 4\;\det[C_0^2]= 0;
\label{eq:monometr1}
\end{equation}
\begin{equation}
C2:\qquad 2\; \tr\left\{ \Omega^2\; C_0^2 \right\} - \tr[\Omega^2]\tr[C_0^2]= 0;
 \label{eq:monometr2}
\end{equation}
to see what they tell us. 
The first of these conditions is equivalent to the
statement that the $2\times2$ matrix $C_0^2$ has two
identical eigenvalues. But since $C_0^2$ is symmetric this then implies 
$C_0^2 = c_0^2 \; \mathbf{I}$, 
 in which case the second condition is automatically satisfied. (In contrast, condition $C2$ does not automatically imply condition $C1$.)
Indeed  if $C_0^2 = c_0^2 \; \mathbf{I}$,
then it is easy to see that (in order to make $C_0^2$ diagonal)
$\tilde U_{AB}=0$,
and furthermore that
\begin{equation}
{\tilde U_{AA} \;\rho_{A0}\over m_A} = c_0^2 = {\tilde
  U_{BB} \;\rho_{B0}\over m_B}.
  \label{eq:c0ft}
\end{equation}
Note that we can now solve
for $\lambda$ to get
\begin{equation}
\lambda = -2 \sqrt{\rho_{A0}\;\rho_{B0}} \; U_{AB},
\label{eq:lbd}
\end{equation}
whence
\begin{equation}
c_0^2 = {U_{AA\;}\rho_{A0}+U_{AB}\; \rho_{B0}\over m_A} =
{U_{BB\;}\rho_{B0}+U_{AB}\; \rho_{A0}\over m_B},
\end{equation} 
and
\begin{equation}
\fl
\omega_{II}^2 = {4\rho_{A0}\rho_{B0}U_{AB}\over \hbar^2}  \left\{
U_{AA} + U_{BB} - 2U_{AB} + U_{AB}  \left[
  \sqrt{\rho_{A0}\over \rho_{B0}} +\sqrt{\rho_{B0}\over
    \rho_{A0}}\right]^2 \right\}.
\label{eq:m-diag}
\end{equation}
Note that (\ref{eq:m-diag}) is equivalent to
(\ref{eq:m-ndiag}) with (\ref{eq:lbd}) enforced. But this then implies
\begin{equation}
\label{E:m2}
\omega_{II}^2 = {4\rho_{A0}\rho_{B0}U_{AB}\over \hbar^2}  \left\{
U_{AA} + U_{BB} + U_{AB}  \left[
  {\rho_{A0}\over \rho_{B0}} +{\rho_{B0}\over
    \rho_{A0}}\right] \right\}.
\label{eq:m-diag2}
\end{equation}
\paragraph{Interpretation:} Condition $C2$ forces the two low-momentum ``propagation speeds'' to be the same, that is, it forces the two $O(k^2)$ coefficients to be equal. Condition $C1$ is the stronger statement that there is no $O(k^4)$ (or higher order) distortion to the relativistic dispersion relation. 

%--------------------------------------------------------------
\section{Beyond the hydrodynamical approximation}%
\label{sec:qp}%
%--------------------------------------------------------------

At this point we want to consider the deviations from the
previous analogue for special relativity. To do so we have to reintroduce
the linearized quantum potential $\hat Q_{X1}$ in our equation
(\ref{phaseequation2}) given that, at high frequencies, it is this term that is inducing deviations from the superfluid
regime for the BEC system.

%-----------------------------------------------------------
\subsection{Fresnel equation}
%-----------------------------------------------------------

Our starting point is equation (\ref{fresnel}) which we
rewrite here for convenience:
\begin{equation}
\omega^2 {\tilde{\theta}}  = 
\left\{
\sqrt{\Xi+X\; k^2} \;\; [D\; k^2+\Lambda]\;\; \sqrt{\Xi+X\;k^2} 
\right\}\;  \tilde{\theta}  \equiv H(k^2) \; \tilde{\theta}.
   \label{eq:new-disp-rel}
\end{equation}
This leads to the Fresnel equation
\begin{equation}
\det\{ \omega^2 \;\mathbf{I} - H(k^2) \} =0.
\end{equation}
That is
\begin{equation}
\omega^4 - \omega^2 \; \hbox{tr}[H(k^2)] + \det[H(k^2)] =
0,
\end{equation}
whence
\begin{equation}
\omega^2 = { \hbox{tr}[H(k^2)] \pm \sqrt{
    \hbox{tr}[H(k^2)]^2 - 4\;  \det[H(k^2)] }\over 2},
\label{eq:tot-disp-rel}
\end{equation}
which is now of the form
\begin{equation}
\omega^2 = [\hbox{quartic}_1] \pm \sqrt{[ \hbox{octic} ]}.
\end{equation}
And we can now proceed with the same sort of analysis as in the hydrodynamical case.

%-----------------------------------------------------------
\subsection{Masses}
%-----------------------------------------------------------

The ``masses'', defined as the zero momentum oscillation frequencies,
are again easy to identify. Just note that
\begin{equation}
H(k^2\to0) = \Omega^2,
\end{equation}
and using the fact that $\det(\Omega^2)=0$ one again obtains
\begin{equation}
\omega^2(k\to 0) = \{0,\tr[\Omega^2]\}.
\end{equation}
So even taking the quantum potential into account we completely
recover our previous results  (\ref{eq:masses}), (\ref{eq:masses2}), and (\ref{eq:m-ndiag}). Of course
this could have been predicted in advance by just
noticing that the $k$-independent term in the Fresnel
equation is exactly the same mass matrix $\Omega^2=\Xi^{1/2}\;\Lambda\;\Xi^{1/2}$ as was present in
the hydrodynamical limit. (That is, the quantum potential term
$\tilde{X}$ does not influence the masses.)

%------------------------------------------------------------------
\subsection{Dispersion relations}
%------------------------------------------------------------------

Let us start again from the general result we obtained for
the dispersion relation (\ref{eq:tot-disp-rel}).
Differently from the previous case, when the hydrodynamic
approximation held, we now have that the discriminant  of
(\ref{eq:tot-disp-rel}) generically can be an eighth-order polynomial
in $k$.  In this case we cannot hope to recover an exact analogue of special
relativity, but instead can at best hope to obtain dispersion relations with vanishing or
{\em suppressed} deviations from special relativity at low $k$; 
possibly with large deviations from special relativity at high momenta. From the form of our
equation it is clear that the Lorentz violation suppression should be somehow associated with the
masses of the atoms $m_{A/B}$. Indeed we will use the underlying atomic masses to define our ``Lorentz breaking scale'', which we shall then assume can be identified with the ``quantum gravity scale''.
The exact form and
relative strengths of the higher-order terms  will be
controlled by tuning the 2--BEC system and will
eventually decide the manifestation (or not) of the
naturalness problem and of the universality issue.

Our approach will again consist of considering derivatives of (\ref{eq:tot-disp-rel}) in growing even powers of $k^2$ (recall that odd powers of $k$ are excluded by the parity invariance of the system) and then setting $k\to0$. We shall compute only the coefficients up to order $k^4$ as by simple dimensional arguments one can expect any higher order term will be further suppressed with respect to the $k^4$ one.

We can greatly simplify our calculations if before performing our analysis we rearrange  our problem in the following way. First of all note that by the cyclic properties of trace
\begin{eqnarray}
\fl
\tr[H(k^2)] &=& \tr[ (Dk^2+\Lambda)\;(\Xi+k^2X)] 
\\
\fl
&=& \tr[ \Lambda \Xi + k^2(D\Xi+\Lambda X) + (k^2)^2 D X]
\\
\fl
&=& \tr[  \Xi^{1/2} \Lambda \Xi^{1/2} + k^2(\Xi^{1/2} D \Xi^{1/2} 
+ X^{1/2}\Lambda X^{1/2})
+ (k^2)^2 X^{1/2} D X^{1/2}].
\end{eqnarray}
We can now define symmetric matrices
\begin{equation}
\Omega^2 =   \Xi^{1/2} \Lambda \Xi^{1/2}; 
\end{equation}
\begin{equation}
C_0^2 = \Xi^{1/2} D \Xi^{1/2}; \qquad
\Delta C^2 =  X^{1/2}\Lambda X^{1/2};
\end{equation}
\begin{equation}
 C^2 = C_0^2 + \Delta C^2 = \Xi^{1/2} D \Xi^{1/2} + X^{1/2}\Lambda X^{1/2}; 
\end{equation}
\begin{equation}
Z^2 =  2 X^{1/2} D X^{1/2} = {\hbar^2\over 2} M^{-2}.
\end{equation}
With all these definitions we can then write
\begin{equation}
\tr[H(k^2)] = \tr\left[ \Omega^2 + k^2 (C_0^2+\Delta C^2) + {1\over2} (k^2)^2 Z^2 \right],
\end{equation}
where everything has been done inside the trace. If we now define
\begin{equation}
H_s(k^2) = \Omega^2 + k^2 (C_0^2+\Delta C^2) + {1\over2} (k^2)^2 Z^2,
\end{equation}
then $H_s(k^2)$ is by definition both polynomial and symmetric and satisfies
\begin{equation}
\tr[H(k^2)] = \tr[H_s(k^2)],
\end{equation}
while in contrast, 
\begin{equation}
\det[H(k^2)] \neq \det[H_s(k^2)].
\end{equation}
But then
\begin{equation}
\omega^2 = {1\over2}\left[ \tr[H_s(k^2)] \pm \sqrt{ \tr[H_s(k^2)]^2-4\det[H(k^2)]}\right].
\end{equation}
Whence
\def\d{{\mathrm{d}}}
\begin{equation}
{\d\omega^2\over\d k^2} = {1\over2}\left[\tr[H_s'(k^2)] \pm 
{
\tr[H_s(k^2)]\tr[H_s'(k^2)] - 2\det'[H(k^2)] \over
\sqrt{ \tr[H_s(k^2)]^2- 4\det[H(k^2)] }
}\right],
\end{equation}
and at $k=0$
\begin{equation}
\left.{\d\omega^2\over\d k^2}\right|_{k\to0} = {1\over2}\left[\tr[C^2] \pm 
{
\tr[\Omega^2]\tr[C^2] - 2\det'[H(k^2)]_{k\to0} \over
 \tr[\Omega^2] 
}\right].
\end{equation}
But now let us consider 
\begin{eqnarray}
\det[H(k^2)] &=& \det[  (Dk^2+\Lambda)\;(\Xi+k^2X) ] 
\\
&=& 
\det[Dk^2+\Lambda]\;\det[\Xi+k^2X] 
\\
&=&
\det[ \Xi^{1/2} (Dk^2+\Lambda) \Xi^{1/2}] \;
\det[ I + k^2 \Xi^{-1/2} X \Xi^{-1/2} ] 
\end{eqnarray}
where we have repeatedly used properties of the determinant.
Furthermore
\begin{eqnarray}
 \det[ I + k^2 \Xi^{-1/2} X \Xi^{-1/2} ] &=& \det[I+k^2 \Xi^{-1} X] 
 \\
&=& \det[I+k^2 X^{1/2} \Xi X^{1/2}] 
 \\
 &=& \det[I+k^2 Y^2/2],
\end{eqnarray}
so that we have
\begin{equation}
\det[H(k^2)] = \det[\Omega^2+C_0^2 k^2] \; \det[I + k^2 Y^2/2].
\end{equation}
Note the the matrix $Y^2$ is the ``healing length matrix'' we had previously defined, and that the net result of this analysis is that the full determinant is the product of the determinant previously found in the hydrodynamic limit  with a factor that depends on the product of wavenumber and healing length.  

But now, given our formula \eref{E:A-2-matrices} for the determinant, we see
\begin{eqnarray}
{\det}'[H(k^2)] &=& (-\tr(\Omega^2\bar C_0^2)+ 2 k^2 \det[C_0^2] ) \; \det[I + k^2 Y^2/2] 
\nonumber
\\
&&
+  \det[\Omega^2+C_0^2 k^2] \; (-\tr[\bar Y^2] +  k^2 \det[Y^2])/2,
\end{eqnarray}
whence
\begin{equation}
{\det}'[H(k^2)]_{k\to0} = -\tr(\Omega^2\bar C_0^2 ),
\end{equation}
and so
\begin{equation}
\left.{\d\omega^2\over\d k^2}\right|_{k\to0} = {1\over2}\left[\tr[C^2] \pm 
{
\tr[\Omega^2]\tr[C^2] + 2 \tr(\Omega^2\bar C_0^2 ) \over
 \tr[\Omega^2] 
}\right].
\end{equation}
That is:
\begin{equation}
\label{eq: varpi2}
\left.{\d\omega^2\over\d k^2}\right|_{k\to0} = 
{1\over2}\left[\tr[C^2] 
\pm 
\left\{ \tr[C^2] + 2{ \tr(\Omega^2\bar C_0^2 ) \over \tr[\Omega^2] }
\right\}
\right].
\end{equation}
Note that all the relevant matrices have been carefully symmetrized. Also note the important distinction between $C_0^2$ and $C^2$. 
Now define
\begin{equation}
c^2 = {1\over2}\tr[C^2],
\end{equation}
then
\begin{equation}
\left.{\d\omega^2\over\d k^2}\right|_{k\to0} = c^2 (1\pm \eta_2),
\end{equation}
with
\begin{equation}
\eta_2 
= \left\{ { \tr[C^2]\tr[\Omega^2] + 2 \tr(\Omega^2\bar C_0^2 ) \over \tr[\Omega^2] \tr[C^2]}
\right\}
 = \left\{ 1 + {\tr(\Omega^2\bar C_0^2 ) \over \omega_{II}^2 \; c^2}
\right\}.
\end{equation}

Similarly, consider the second derivative: 
\begin{eqnarray}
\fl
{\d^2\omega^2\over\d(k^2)^2} &=&  {1\over2}\Bigg[\tr[H_s''(k^2)] 
\nonumber
\\
\fl
&&
\pm 
{
\tr[H_s(k^2)]\tr[H_s''(k^2)] + \tr[H_s'(k^2)]\tr[H_s'(k^2)]- 2\det''[H(k^2)] \over
\sqrt{ \tr[H_s(k^2)]^2- 4\det[H(k^2)] }
}
\nonumber
\\
\fl
&&
\mp 
{
( \tr[H_s(k^2)]\tr[H_s'(k^2)] - 2\det'[H(k^2)] )^2\over
(\tr[H_s(k^2)]^2- 4\det[H(k^2)] )^{3/2}
}
\Bigg],
\end{eqnarray}
whence
\begin{eqnarray}
\left.{\d^2\omega^2\over\d(k^2)^2}\right|_{k\to0} &=&  {1\over2}\Bigg[\tr[Z^2] 
\pm 
{
\tr[\Omega^2]\tr[Z^2] + \tr[C^2]^2- 2\det''[H(k^2)]_{k\to0} \over
\tr[\Omega^2]
}
\nonumber
\\
&&
\mp 
{
( \tr[\Omega^2]\tr[C^2] - 2\det'[H(k^2)]_{k\to0} )^2\over
\tr[\Omega^2]^3
}
\Bigg].
\end{eqnarray}
The last term above can be related to $\d\omega^2/\d k^2$, while the determinant piece is evaluated using
\begin{eqnarray}
{\det}''[H(k^2)] &=& (2\det[C_0^2] ) \; \det[I + k^2 Y^2/2] 
\\
&&
+ (-\tr(\Omega^2\bar C_0^2)+ 2 k^2 \det[C_0^2] ) \;  (-\tr[\bar Y^2] +  k^2 \det[Y^2])/2
\nonumber
\\
&&
+  \det[\Omega^2+C_0^2 k^2] \; ( \det[Y^2]/2)
\nonumber
\\
&&
+   (-\tr(\Omega^2\bar C_0^2)+ 2 k^2 \det[C_0^2] )  \; (-\tr[\bar Y^2] +  k^2 \det[Y^2])/2.
\nonumber
\end{eqnarray}
Therefore
\begin{eqnarray}
{\det}''[H(k^2)]_{k\to0} &=& (2\det[C_0^2] ) 
\nonumber
\\
&&
+ (-\tr(\Omega^2\bar C_0^2) ) \;  (-\tr[\bar Y^2] )/2
+  \det[\Omega^2] \; (\det[Y^2])/2
\nonumber
\\
&&
+   (-\tr(\Omega^2\bar C_0^2))  \; (-\tr[\bar Y^2])/2.
\end{eqnarray}
That is, (recalling $\tr[\bar A] = - \tr[A]$),
\begin{equation}
{\det}''[H(k^2)]_{k\to0} = (2\det[C_0^2] ) 
-(\tr(\Omega^2\bar C_0^2) ) \;  (\tr[Y^2] ),
\end{equation}
or
\begin{equation}
{\det}''[H(k^2)]_{k\to0} = -\tr[C_0^2 \bar C_0^2]
- \tr[\Omega^2\bar C_0^2] \;  \tr[Y^2] .
\end{equation}
Now assembling all the pieces
\begin{eqnarray}
\fl
\left.{\d^2\omega^2\over\d(k^2)^2}\right|_{k\to0} &=  {1\over2}\Bigg[\tr[Z] 
\pm 
{
\tr[\Omega^2]\tr[Z^2] + \tr[C^2]^2+ 2  \tr[C_0^2 \bar C_0^2] 
+2 (\tr(\Omega^2\bar C_0^2) ) \;  (\tr[Y^2] )
\over
\tr[\Omega^2]
}
\nonumber
\\
&
\mp 
{
( \tr[\Omega^2]\tr[C^2] + 2\tr[\Omega^2 \bar C_0^2] )^2\over
\tr[\Omega^2]^3
}
\Bigg].
\end{eqnarray}
That is
\begin{eqnarray}
\fl
\left.{\d^2\omega^2\over\d(k^2)^2}\right|_{k\to0} &=  {1\over2}\Bigg[\tr[Z] 
\pm 
{
\tr[\Omega^2]\tr[Z^2] + \tr[C^2]^2+ 2  \tr[C_0^2 \bar C_0^2] 
+2 (\tr(\Omega^2\bar C_0^2) ) \;  (\tr[Y^2] )
\over
\tr[\Omega^2]
}
\nonumber
\\
&
\qquad
\mp 
{\tr[C^2]^2\over\tr[\Omega^2]} \eta_2^2\Bigg],
\end{eqnarray}
and so
\begin{eqnarray}
\fl \left.{\d^2\omega^2\over\d(k^2)^2}\right|_{k\to0} =  &{\textstyle 1\over \textstyle 2}&\Bigg[
\tr[Z^2]  \pm \tr[Z^2] 
\pm 2 \frac{\tr[\Omega^2\bar{C}^2_0]}{\tr[\Omega^2]}\tr[Y^2]
\pm {\tr[C^2]^2- 4\det[C^2_0] \over \tr[\Omega^2] }\nonumber\\
&\mp& 
{\tr[C^2]^2\over\tr[\Omega^2]} \eta_2^2
\Bigg].\label{eq: varpi4}
\end{eqnarray}

With the above formula we have completed our derivation of the  lowest-order terms of the generic dispersion relation of a coupled 2-BEC system --- including the terms introduced by the quantum potential at high wavenumber --- up to terms of order $k^4$.
From the above formula it is clear that we do not generically have Lorentz invariance in this system: Lorentz violations arise both due to mode-mixing interactions (an effect which can persist in the hydrodynamic limit where $Z\to0$ and $Y\to0$) and to the presence of the quantum potential (signalled by $Z\neq0$ and $Y\neq0$). 
While the mode-mixing effects are relevant at all energies the latter effect characterizes the discrete structure of the effective spacetime at high energies. It is in this sense that the quantum potential determines the analogue of quantum gravity effects in our 2-BEC system. 
%

%---------------------------------------------
\section{The relevance for quantum gravity phenomenology}
\label{sec:qgp}
%---------------------------------------------

Following this physical insight we can now easily identify a regime that is potentially relevant for simulating the typical ans\"atze of quantum gravity phenomenology. 
We demand that any violation of Lorentz invariance present should be due to the microscopic structure of the effective spacetime. This implies that one has to tune the system in order to cancel exactly all those violations of Lorentz invariance which are solely due to mode-mixing interactions in the hydrodynamic limit. 

We basically follow the guiding idea that a good analogue of quantum-gravity-induced Lorentz violations should be characterized only by the ultraviolet physics of the effective spacetime. In the system at hand the ultraviolet physics is indeed characterized by the quantum potential,  whereas possible violations of the Lorentz invariance in the hydrodynamical limit are low energy effects, even though they have their origin in the microscopic interactions. We therefore start by investigating the scenario in which the system is tuned in such a way that no violations of Lorentz invariance are present in the hydrodynamic limit. This leads us to again enforce the conditions $C1$ and $C2$ which corresponded to ``mono-metricity" in the hydrodynamic limit. 

In this case (\ref{eq: varpi2}) and  (\ref{eq: varpi4}) take respectively the form
\begin{eqnarray}
\left.{\d\omega^2\over\d k^2}\right|_{k\to0} &=& 
{1\over2}\left[\tr[C_0^2] +(1\pm1)\,\tr[\Delta C^2]\right]
= c_0^2 + {1\pm1\over2}\tr[\Delta C^2],\label{eq:varpi2b}
\end{eqnarray}
and
\begin{eqnarray}
\left.{\d^2\omega^2\over\d(k^2)^2}\right|_{k\to0} &=
{\tr[Z^2]  \pm \tr[Z^2] \over 2}\mp\tr[C^2_0]\tr[Y^2]
\nonumber
\\
& 
\pm{1\over 2} {\tr[\Delta C^2]^2+2\tr[C^2_0]\tr[\Delta C^2] \over \tr[\Omega^2] }
\mp {1\over 2}
{
\tr[\Delta C^2]^2\over
\tr[\Omega^2]
}
\nonumber
\\
&=
{\tr[Z^2]  \pm \tr[Z^2] \over 2}\pm\tr[C^2_0]\left(-\tr[Y^2]+
{\tr[\Delta C^2] \over \tr[\Omega^2] }\right).
\label{eq:varpi4b}
\end{eqnarray}

Recall (see section \ref{S:C1C2}) that the first of the physical conditions $C1$ is equivalent to the statement that the $2\times2$ matrix $C_0^2$ has two
identical eigenvalues. But since $C_0^2$ is symmetric this then implies $C_0^2 = c_0^2 \; \mathbf{I}$,  in which case the second condition is automatically satisfied.
This also leads to the useful facts
\begin{eqnarray}
&&\tilde{U}_{AB}=0 \quad \Longrightarrow \quad \lambda = -2 \sqrt{\rho_{A0}\;\rho_{B0}} \; U_{AB}
\label{eq:lambdaUAB} ;\\
&& c_0^2 ={\tilde U_{AA} \;\rho_{A0}\over m_A} =  {\tilde
  U_{BB} \;\rho_{B0}\over m_B}.
    \label{eq:c0ft2}
\end{eqnarray}
Now that we have the fine tuning condition for the laser coupling we can compute the magnitude of the effective mass of the massive phonon and determine the values of the Lorentz violation coefficients. In particular we shall start checking that this regime 
allows for a real positive effective mass as needed for a suitable analogue model of quantum gravity phenomenology.

%---------------------------------------
\subsection{Effective mass:}
%---------------------------------------

Remember that  the definition of $m_{II}$ reads 
\begin{equation}
m_{II}^2 = \hbar^2 \omega_{II}^2/c_0^4.
\label{eq:m2-c1}
\end{equation}
Using equation~(\ref{eq:lambdaUAB}) and equation~(\ref{eq:c0ft2})
we can rewrite $c^2_0$ in the following form
\begin{equation}
\fl
c_0^2 = [m_B \rho_{A0} U_{AA} + m_A \rho_{B0} U_{BB} 
         + U_{AB} (\rho_{A0} m_A + \rho_{B0} m_B)] /
         (2 m_A m_B).
         \label{eq:c0av}
\end{equation}
Similarly equation~(\ref{eq:lambdaUAB}) and equation~(\ref{eq:c0ft2}) when inserted  in  equation (\ref{E:m2}) give
\begin{equation}
\omega_{II}^2 = \frac{4 U_{AB} (\rho_{A0} m_B + \rho_{B0} m_A) c_0^2}{ \hbar^2}.
\label{eq:om2}
\end{equation}
We can now estimate $m_{II}$ by simply inserting the above expressions in equation~(\ref{eq:m2-c1}) so that 
\begin{equation}
m_{II}^2 = {
8 U_{AB} (\rho_{A0} m_A+\rho_{B0} m_B) m_A m_B 
\over
          [m_B \rho_{A0} U_{AA} + m_A \rho_{B0} U_{BB} 
         + U_{AB} (\rho_{A0} m_A + \rho_{B0} m_B)]
}.
\end{equation}
This formula is still a little clumsy but a great deal can be understood by doing the physically reasonable approximation $m_A \approx m_B=m$  and $\rho_A \approx \rho_B$. In fact in this case one obtains
\begin{equation}
m_{II}^2 \approx m^2  \; {
8 U_{AB}
\over [U_{AA}+2U_{AB}+U_{BB}]
}.
\end{equation}
This formula now shows clearly that, as long as the mixing term $U_{AB}$ is small compared to the ``direct" scattering $U_{AA}+U_{BB}$, the mass of the heavy phonon will be ``small" compared to the mass of the atoms.  Though experimental realizability of the system is not the primary focus of the current article, we point out that there is no obstruction in principle to tuning a 2-BEC system into a regime where $|U_{AB}| \ll |U_{AA}+U_{BB}|$. 
For the purposes of this paper it is sufficient that a small effective phonon mass (small compared to the atomic masses which set the analogue quantum gravity scale) is obtainable for some arrangement of the microscopic parameters.
We can now look separately at the coefficients of the quadratic and quartic Lorentz violations and then compare their relative strength in order to see if a situation like that envisaged by discussions of the naturalness problem is actually realized.

%-----------------------------------------------------------------
\subsection{Coefficient of the quadratic deviation:}
%-----------------------------------------------------------------

One can easily see from (\ref{eq:varpi2b}) that the $\varpi_2$ coefficients for this case take the form
\begin{eqnarray}
\varpi_{2,I} &=& 0;\\
\varpi_{2,II} &=&\tr[ \Delta C^2]=\tr[X^{1/2} \Lambda X^{1/2}]=\tr[X\Lambda]\nonumber
\\
&=& -\frac{1}{2}\frac{\lambda}{m_Am_B}\left( 
\frac{m_A\rho_{A0}+m_B\rho_{B0}}
{\sqrt{\rho_{A0}\rho_{B0}}} \right).
\end{eqnarray}

%%%%%%%
So if we insert the fine tuning condition for $\lambda$, equation~(\ref{eq:lambdaUAB}), we get 
\begin{eqnarray}
\eta_{2,II} &=&\frac{\varpi_{2,II}}{c^2_0}=\frac{U_{AB}\left( 
m_A\rho_{A0}+m_B\rho_{B0}\right)
}{m_Am_Bc_0^2}.
\label{E:eta_2II}
\end{eqnarray}
Remarkably we can now cast this coefficient in a much more suggestive form by expressing the coupling $U_{AB}$ in terms of the mass of the massive quasi-particle $m_{II}^2$.
In order to do this we  start from equation (\ref{eq:om2}) and note that it 
enables us to express $U_{AB}$ in (\ref{E:eta_2II}) in terms of $\omega_{II}^2$,  thereby  obtaining
\begin{equation}
\eta_{2,II} =\frac{\hbar^2}{4 c^4_0} \; 
\frac{\rho_{A0} m_A + \rho_{B0} m_B}{\rho_{A0} m_B + \rho_{B0} m_A} \;
\frac{\omega_{II}^2 }{m_Am_B}.
\end{equation}
Now it is easy to see that
\begin{equation}
\frac{\rho_{A0} m_A + \rho_{B0} m_B}{\rho_{A0} m_B + \rho_{B0} m_A}
\approx \mathcal{O} (1),
\end{equation}
and that this factor is identically unity if either $m_A = m_B$ or $\rho_{A0} = \rho_{B0}$.
All together we are left with
\begin{equation} 
\eta_{2,II}  = \bar{\eta} \left(
\frac{m_{II} }{\sqrt{m_Am_B}} \right)^2, 
\label{eq:eta2qfin}
\end{equation}
where $\bar{\eta}$ is a dimensionless coefficient of order one.\\

The product in the denominator of the above expression can be interpreted as the geometric mean of the fundamental bosons masses $m_A$ and $m_B$. These are mass scales associated with the microphysics of the condensate --- in analogy with our experience with a 1-BEC system where the ``quantum gravity scale" $K$ in equation (\ref{bogo-disp}) is set by the mass of the BEC atoms. It is then natural to define  an analogue of the scale of the breakdown of Lorentz invariance as $M_{\rm eff}=\sqrt{m_Am_B}$.   (Indeed this ``analogue Lorentz breaking scale'' will  typically do double duty as an ``analogue Planck mass".)

Using this physical insight it should be clear that equation~(\ref{eq:eta2qfin}) effectively says
\begin{equation}
\eta_{2,II}\approx\left(\frac{m_{II}}{M_{\rm eff}}\right)^2,
\label{eq:eta2fin}
\end{equation}
which, given that $m_I=0$, we are naturally lead to generalize to
\begin{equation}
\fl
\eta_{2,X}\approx\left(\frac{m_X}{M_{\rm eff}}\right)^2= 
\left( {\hbox{mass scale of quasiparticle}
\over
\hbox{effective Planck scale}}\right)^2; \qquad X=I,II.
\label{eq:eta2finbb}
\end{equation}
The above relation is exactly the sort of dimensionless ratio $(\mu/M)^\sigma$ that has been very often \emph{conjectured} in the literature on quantum gravity phenomenology in order to explain the strong observational constraints on  Lorentz violations at the lowest orders. (See discussion in the introduction.)
Does this now imply that this particular regime of our 2-BEC system will also show an analogue version of the naturalness problem? In order to answer this question we need to find the dimensionless coefficient for the quartic deviations, $\eta_4$, and check if it will or won't itself be suppressed by some power of the small ratio $m_{II}/M_{\rm eff}$.

%-----------------------------------------------------------------
\subsection{Coefficients of the quartic deviation:}
%------------------------------------------------------------------

Let us now consider  the coefficients of the quartic term presented in equation~(\ref{eq:varpi4b}). 
For the various terms appearing in (\ref{eq:varpi4b}) we get
\begin{equation}
\tr[Z^2]=2\tr[DX]=
\frac{\hbar^2}{2}\left(\frac{m^2_A+m^2_B}{m^2_A m^2_B}\right);
\end{equation}
\begin{equation}
\tr[\Delta C^2]=\tr[X \Lambda]=
-\frac{\lambda}{2}\frac{m_A\rho_{A0}+m_B\rho_{B0}}
{m_A m_B\sqrt{\rho_{A0}\rho_{B0}}}
=
U_{AB}\frac{m_A\rho_{A0}+m_B\rho_{B0}}{m_A m_B};
\end{equation}
\begin{equation}
\tr[Y^2]=2\tr[X\Xi^{-1}]=\frac{\hbar^2}{2}
\frac{\rho_{A0}m_{A}\tilde{U}_{AA}+\rho_{B0}m_{B}\tilde{U}_{BB}}
{\rho_{A0}m_{A}\rho_{B0} m_{B} \tilde{U}_{AA}\tilde{U}_{BB}};
\end{equation}
where in the last expression we have used the fact that in the current scenario $\tilde{U}_{AB}=0$.
Now by definition 
\begin{equation}
\eta_4= ({M_\mathrm{eff}^2}/{\hbar^2})\; \varpi_4 = {1\over2} ({M_\mathrm{eff}^2}/{\hbar^2}) \left[{\d^2\omega^2\over (\d k^2)^2}\right]_{k=0}
\end{equation}
 is the dimensionless coefficient in front of the $k^4$. So
\begin{eqnarray}
\eta_4 &=& \frac{M_\mathrm{eff}^2}{2\hbar^2} \left[ {\tr[Z^2]  \pm \tr[Z^2] \over 2}\pm\tr[C^2_0]\left(-{\tr[Y^2]\over2}+
{\tr[\Delta C^2] \over \tr[\Omega^2] }\right)\right]\\
&=& 
\frac{ M_\mathrm{eff}^2\; c_0^2}{\hbar^2} 
\left[ 
{\tr[Z^2]  \pm \tr[Z^2] \over 2\tr[C^2_0]}\pm\left(-{\tr[Y^2]\over2}+
{\tr[\Delta C^2] \over \tr[\Omega^2] }\right)\right].
\end{eqnarray}
Whence 
\begin{eqnarray}
\eta_{4,I} &=& \frac{ M_\mathrm{eff}^2\;c_0^2}{\hbar^2}
\left[ {\tr[Z^2] \over \tr[C^2_0]}+\left(-{\tr[Y^2]\over2}+
{\tr[\Delta C^2] \over \tr[\Omega^2] }\right)\right]
\\
\eta_{4,II} &=& \frac{ M_\mathrm{eff}^2\;c_0^2}{\hbar^2}
\left[ \left({\tr[Y^2]\over2}-
{\tr[\Delta C^2] \over \tr[\Omega^2] }\right)\right].
\end{eqnarray}
Let us compute the two relevant terms separately:
\begin{eqnarray}
{\tr[Z^2] \over \tr[C^2_0]} &=&
\frac{\hbar^2}{4c_0^2}\left(\frac{m^2_A+m^2_B}{m^2_A m^2_B}\right)=
\frac{\hbar^2}{4 c_0^2 M_\mathrm{eff}^2}
\left(\frac{m^2_A+m^2_B}{m_A m_B}\right);
\\
\fl
-\tr[Y^2]/2+
{\tr[\Delta C^2] \over \tr[\Omega^2] } &=& 
-\frac{\hbar^2}{4 M_\mathrm{eff}^2}
\left[
\frac{\rho_{A0}m_{A}\tilde{U}_{AA}^2+\rho_{B0}m_{B}\tilde{U}_{BB}^2}
{\rho_{A0}\rho_{B0}\tilde{U}_{AA}\tilde{U}_{BB}
\left(\tilde{U}_{AA}+\tilde{U}_{BB}\right)}
\right]
\nonumber\\
&=& 
-\frac{\hbar^2}{4 M_\mathrm{eff}^2\;c_0^2}
\left[
\frac{m^2_{A}\tilde{U}_{AA}+m^2_{B}\tilde{U}_{BB}}
{m_{A}m_{B}\left(\tilde{U}_{AA}+\tilde{U}_{BB}\right)}
\right]
\end{eqnarray}
where we have used $\rho_{X0}\tilde{U}_{XX}=m_{X}c^2_0$ for $X=A,B$ as in equation~(\ref{eq:c0ft2}).
 Note that the quantity in square brackets in the last line is dimensionless.
So in the end:
\begin{eqnarray}
\fl \eta_{4,I} &=&  \frac{1}{4}\left[\left(\frac{m^2_A+m^2_B}{m_A m_B}\right)-\frac{m^2_{A}\tilde{U}_{AA}+m^2_{B}\tilde{U}_{BB}}
{m_{A}m_{B}\left(\tilde{U}_{AA}+\tilde{U}_{BB}\right)}
\right]=\frac{1}{4}
\left[
\frac{m^2_{A}\tilde{U}_{BB}+m^2_{B}\tilde{U}_{AA}}
{m_{A}m_{B}\left(\tilde{U}_{AA}+\tilde{U}_{BB}\right)}\right];
\label{eq:eta4I}\\
\fl \eta_{4,II} &=& \frac{1}{4}
\left[
\frac{m^2_{A}\tilde{U}_{AA}+m^2_{B}\tilde{U}_{BB}}
{m_{A}m_{B}\left(\tilde{U}_{AA}+\tilde{U}_{BB}\right)}\right].
\label{eq:eta4II}
\end{eqnarray}
\paragraph{Note:} In the special case $m_A=m_B$ we recover identical quartic deviations $\eta_{4,I}=\eta_{4,II}=1/4$, indicating in this special situation a ``universal'' deviation from Lorentz invariance. Indeed we also obtain $\eta_{4,I}=\eta_{4,II}$ if we demand $\tilde{U}_{AA}=\tilde{U}_{BB}$, even without fixing $m_A=m_B$.

%--------------------------------------------------------------------
\subsection{Avoidance of the naturalness problem:}
%--------------------------------------------------------------------

We can  now ask ourselves if there is, or is not, a naturalness problem present in our system. Are the dimensionless coefficients $\eta_{4,I/II}$ suppressed below their naive values by some small ratio involving $M_{\rm eff}=\sqrt{m_Am_B}$~? Or are these ratios unsupressed? 
Indeed at first sight it might seem that further supression is the case, since  the square of  the ``effective Planck scale" seems to appear in the denominator of both the coefficients  (\ref{eq:eta4I}) and (\ref{eq:eta4II}). However, the squares of the atomic masses also appear in the numerator, rendering both coefficients of order unity.

It is perhaps easier  to see this once the dependence of  (\ref{eq:eta4I}) and (\ref{eq:eta4II}) on the effective coupling $\tilde{U}$ is removed. We again use the substitution $\tilde{U}_{XX}=m_{X}c^2_0/\rho_{X0}$ for $X=A,B$, so obtaining:
\begin{eqnarray}
\eta_{4,I} &=&  \frac{1}{4}
\left[
\frac{m_{A}\rho_{A0}+m_{B}\rho_{B0}}{m_{A}\rho_{B0}+m_{B}\rho_{A0}}
\right];
\label{eq:eta4Ifin}\\
&&\nonumber\\
\eta_{4,II} &=& \frac{1}{4}
\left[
\frac{m_{A}^3 \rho_{B0}+m_{B}^3 \rho_{A0}} 
{
m_A m_B  \;( m_{A}\rho_{B0}+m_{B}\rho_{A0})}
\right].
\label{eq:eta4IIfin}
\end{eqnarray}
From these expressions is clear that the $\eta_{4,I/II}$ coefficients are actually of order unity. 

That is, if our system is set up so that $m_{II}\ll m_{A/B}$ --- which we have seen in this scenario is equivalent to requiring $U_{AB}\ll U_{AA/BB}$ --- no naturalness problem arises as for $p > m_{II}\;c_0$ the higher-order, energy-dependent Lorentz-violating terms ($n\geq 4$) will indeed dominate  over the quadratic Lorentz-violating term.

%-------------------------------
\section{Summary and Discussion}
\label{sec:discuss}
%-------------------------------

Analogue models of gravity have a manifold role which can be summarized in three main points: (1) They can reproduce in a laboratory what we believe are the most important features of QFT in curved spacetimes; (2) they can give us  the possibility of understanding the phenomenology of  condensed matter systems via the body of knowledge developed in semiclassical gravity; finally (3) they can be used as test fields and inspiration for new ideas about the nature and consequences of an emergent spacetime~\cite{abh:emergent, normal, refringence, bimetricity}.

In this paper we have followed the last path by studying an analogue system which allows us to test the conjectures that lie at the base of most of the extant literature on quantum gravity phenomenology --- by building an analogue spacetime exhibiting Planck-suppressed Lorentz violations. 
This analogue model, arising from a coupled 2-BEC system (previously studied in~\cite{VW1, VW2}), reveals itself as an ideal system for reproducing the salient features of the most common ans\"atze for quantum gravity phenomenology.
Excitations in a coupled 2-BEC system result in the analogue kinematics for a massive and a massless scalar field. For a fine tuning in the hydrodynamic limit --- sufficient to describe low-energy excitations --- we recovered perfect Lorentz invariance.

To describe highly energetic modes we modified the theory by including the quantum potential term. This is a quantum correction to the classical mean-field, which is energy-dependent and therefore is no longer negligible at high energy.
We developed several mathematical tools to analyze the dispersion relations arising in this system. We Taylor-expanded the system around $k=0$, and calculated the coefficients for the quadratic and quartic order terms.
We considered first the hydrodynamical approximation (phonons of long wavelength) as this case in a 1-BEC system leads unequivocally to a special relativistic kinematics. In the 2-BEC system we found that only for some specific fine-tuned values of the laser coupling $\lambda$ is a single background relativistic kinematics  (``mono-metricity'')  for the two phonon modes recovered. This is as expected as it is well known that for complex systems mono-metricity is not the only outcome for the propagation of linearized perturbations~\cite{normal, refringence, bimetricity}. 

We then relaxed the approximation of long wavelengths for the 2-BEC system excitations. This allows to consider those Lorentz violations which are really due to the UV physics of the condensate, and which are indeed the source of the standard quartic Lorentz-violating term in the Bogolubov dispersion relation~(\ref{bogo-disp}). By using an eikonal approximation we then studied the situation in which we again enforced the complete suppression of all the Lorentz-violating terms which are not explicitly due to the UV physics, (of course this implies a fine tuning that is identical to that in the hydrodynamic approximation).  

In the analogue spacetime so obtained we found that the issue of {\em universality} 
is fundamentally related to the complexity of the underlying microscopic system. As long as we keep the two atomic masses $m_A$ and $m_B$ distinct we generically have distinct $\eta_4$ coefficients (and the $\eta_2$ coefficients are unequal even the limit $m_A=m_B$). However we can easily recover identical  $\eta_4$ coefficients, for instance,  as soon as we impose identical microphysics for the two BEC systems we couple.
Even more interestingly we saw that due to the presence of the interactions between the two BEC components, the quantum-potential-dependent Lorentz violations not only induce terms at order $k^4$, but also at order $k^2$ (in close analogy with what was predicted in~\cite{Collins} for a generic EFT with higher order Lorentz violations). Remarkably, we find that the $\eta_2$ coefficients are exactly of the form envisaged (within the context of standard quantum gravity phenomenology) in order to be subdominant with respect to the higher-order ones. This implies that our 2-BEC analogue spacetime is an explicit example where the typical dispersion relations used in quantum gravity phenomenology studies are reproduced, and that the {\em naturalness problem} does not arise.

Let us now comment briefly about the nature of this result.
First, it is important to stress that we did not merely perform a ``tree level" calculation in the quasi-particle EFT. The dispersion relations we obtained were computed directly from the true physical {\em microscopic} field theory describing the 2-BEC atomic system, and as such they already consistently take into account all the corrections due to higher loops in the quasi-particle EFT. In this sense the modified dispersion relations we found are those one would expect to observe if an actual experiment is set up, much as the form and the coefficients of the  Bogoliubov dispersion relation are experimentally confirmed by present-day experiments~\cite{Vogels:2002qd}.
Second, we stress that the avoidance of the naturalness problem is not related directly to the fact that we tuned our system so to reproduce special relativistic dispersion relations in the hydrodynamic limit. In fact our conditions for recovering SR at low energies do not \emph{a priori} fix the the $\eta_2$ coefficient,  as its strength after the ``fine tuning" could still be large (even of order one) if the typical mass scale of the massive phonon is not well below the atomic mass scale. Indeed the smallness of the $\eta_2$ coefficient is in this sense  directly related to the mechanism providing a mass to one of the two phonons as we shall discuss at length below.

The question we now want to address is why our model escapes the naive predictions of large Lorentz violations at low energies? 
There is a nice interpretation of this result in terms of ``emergent symmetry''.

We have seen that a non-zero $\lambda$  \emph{simultaneously} produces a non-zero mass term for one of the phonons, \emph{and} a corresponding non-zero LIV at order $k^2$ (single BEC systems have only $k^4$ LIV as described by the Bogoliubov dispersion relation). Let us now imagine  driving $\lambda\to 0$ but keeping  the conditions $C1$ and $C2$ valid at each stage (this requires $U_{AB}\to 0$ as well). In this case one gets an EFT theory which at low energies describes two non-interacting phonons propagating on a common background (in fact $\eta_2\to0$ and $c_I=c_{II}=c_0$). This system posses an $SO(2)$ symmetry corresponding to the invariance under rigid rotations of the doublet formed by the two phonon fields. Recovery of SR at low energies could then be seen as a by-product of imposing this symmetry on the system of two massless phonons. Hence non zero laser coupling $\lambda$ corresponds to a soft breaking of the SO(2) symmetry and of the corresponding Lorentz invariance at low energies. Such violation is then expected to be determined (as usual in EFT) by the ratio of the scale of the symmetry breaking $m_{II}$ and that of the scale originating the LIV in first place $M_{\rm LIV}$. We stress that the SO(2) symmetry is an emergent symmetry as it is not preserved beyond the hydrodynamic limit: the $\eta_4$ coefficients are in general different if $m_A\neq m_B$ so $SO(2)$ it is generically broken at high energies. However this is enough for the protection of the {\em lowest}-order LIV operators. 

Hence this analogue model seems to be telling us that in solving the naturalness problem a possible mechanism could be that of an EFT which in the low-energy limit exhibits an accidental/emergent symmetry, instead of a fundamental one. Note that this is an interesting and original suggestion which goes in the opposite direction with respect to other attempts at solving the naturalness problem by looking at symmetries which are supposed to be exact in the high energy regime (see, \emph{e.g.},  works exploring the role of SUSY, such as~\cite{Pospelov-Nibbelink}, and the related problems with the low-energy symmetry breaking). The lesson that can instead be drawn in this case is that emergent symmetries could be sufficient to minimize the amount of Lorentz violation in the lowest dimension operators of the EFT. 

It is intriguing to think that an interpretation of SUSY  as an accidental symmetry has indeed been considered in recent times~\cite{Luty}, and that this is done at the cost of renouncing attempts to solve the hierarchy problem in the standard way. It might be that in this sense the smallness of the particle physics mass scales with respect to the Planck scale could be directly related to smallness of Lorentz violations in renormalizable operators of the low-energy effective field theory we live in.
We hope to further investigate these issues in a future work.

%-------------------------------------------------------------------
%\clearpage
%--------------------------------------------------------------------
%--------------------------------------------------------------------
\ack

The authors wish to thank David Mattingly, Ted Jacobson and Bei-Lok Hu for illuminating discussions.
This research was partially supported by the Marsden Fund administered by the Royal Society of New Zealand.

%-----------------------------------
\appendix
%-----------------------------------
\section{Some matrix identities}
%-----------------------------------

To simplify the flow of argument in the body of the
paper, here we collect a few basic results on $2\times 2$
matrices that are used in our analysis.

%-----------------------------
\subsection{Determinants}
%-----------------------------

\noindent{\bf Theorem:} For any two $2\times 2$ matrix
$A$:
\begin{equation}
\label{E:A-det2}
\det(A)=  {1\over2} \left\{ \tr[A]^2-\tr[A^2]\right\} . 
\end{equation}
This is best proved by simply noting
\begin{equation}
\det (A) = \lambda_1 \lambda_2 = 
{1\over2}\left[ (\lambda_1 +\lambda_2)^2 - (\lambda_1^2 +\lambda_2^2) \right]
= {1\over2} \left\{ \tr[A]^2-\tr[A^2]\right\} . 
\end{equation}
If we now define $2\times2$ ``trace reversal'' (in a manner reminiscent of standard GR) by
\begin{equation}
\bar A = A - \tr[A]\;\mathbf{I};  \qquad   \bar{\!\!\bar A} = A;
\end{equation}
then this looks even simpler
\begin{equation}
\label{E:A-det23}
\det(A)=  -{1\over2} \tr[A\;\bar A] = \det(\bar A). 
\end{equation}

\noindent
A simple implication is now:\\
\noindent{\bf Theorem:} For any two $2\times 2$ matrices
$A$ and $B$:
\begin{equation}
\label{E:A-2-matrices2}
\det(A+\lambda \; B) = \det(A) + \lambda\;\left\{ \tr[A] \tr[B] - \tr[  A\;B]
\right\}  + \lambda^2\;\det(B).
\end{equation}
which we can also write as
\begin{equation}
\label{E:A-2-matrices}
\det(A+\lambda \; B) = \det(A) - \lambda\; \tr[  A\;\bar B]
+ \lambda^2\;\det(B).
\end{equation}
Note that $\tr[A\;\bar B] = \tr[\bar A\; B]$.

\noindent
By iterating this theorem twice, we can easily see
that:\\
\noindent{\bf Theorem:} For any three   $2\times 2$
matrices  $A$, $B$, and $C$:
\begin{eqnarray}
\label{E:A-3-matrices}
\fl
\det[A+\lambda \; B+\lambda^2\; C] &=& 
\det[A] -
\lambda\;\tr\{ A\;\bar B \}  
+ \lambda^2\;\left[ \det[B] - \tr\{
  A\;\bar C \}   \right]
\nonumber\\ &&\qquad -\lambda^3\; \tr\{
 B\;\bar C \} +\lambda^4 \; \det[C]. 
\end{eqnarray}

%-----------------------------
\subsection{Hamilton--Cayley theorems}
%-----------------------------

\noindent{\bf Theorem:} For any two $2\times 2$ matrix
$A$:
\begin{equation}
A^{-1} = \frac{\tr[A] \;\; \mathbf{I} - A }{\det[A]} = -{\bar A\over\det{[\bar A]}} .
\end{equation}

\noindent{\bf Theorem:} For any two $2\times 2$ matrix
$A$:
\begin{equation}
A^{1/2} = \;\pm \left\{
\frac{A\pm\sqrt{\det A} \;\; \mathbf{I} }{\sqrt{\tr[A]\pm 2  \sqrt{\det A}}}
\right\} .
\end{equation}
%

%-----------------------------------------------
\section*{References}
%-----------------------------------------------

%-----------------------------------

%------------------------------------------

\begin{thebibliography}{69}
%-----------------------------------


\bibitem{LIV}
  D.~Mattingly,
  {\em ``Modern tests of Lorentz invariance,''}
  Living Rev.\ Rel.\  {\bf 8}, 5 (2005)
  [arXiv:gr-qc/0502097].
  %%CITATION = GR-QC 0502097;%%
  
 %--------------------------
\bibitem{Jacobson:2002hd}
T.~Jacobson, S.~Liberati and D.~Mattingly, ``Threshold effects and
Planck scale Lorentz violation: Combined constraints from high
energy astrophysics,'' Phys.\ Rev.\ D {\bf 67}, 124011 (2003)
[arXiv:hep-ph/0209264].
%%CITATION = HEP-PH 0209264;%%
 \\
 D.~Mattingly, T.~Jacobson and S.~Liberati,
  ``Threshold configurations in the presence of Lorentz violating dispersion
  relations,''
  Phys.\ Rev.\ D {\bf 67} (2003) 124012
  [arXiv:hep-ph/0211466].
  %%CITATION = HEP-PH 0211466;%%
  
  
%----------------------------------------------------------------
\bibitem{TeV-QG}

  N.~Arkani-Hamed, S.~Dimopoulos and G.~R.~Dvali,
  {\em ``Phenomenology, astrophysics and cosmology of theories with  sub-millimeter
  dimensions and TeV scale quantum gravity,''}
  Phys.\ Rev.\ D {\bf 59}, 086004 (1999)
  [arXiv:hep-ph/9807344];
  %%CITATION = HEP-PH 9807344;%%
\\
  S.~Dimopoulos and G.~Landsberg,
  {\em ``Black holes at the LHC,''}
  Phys.\ Rev.\ Lett.\  {\bf 87}, 161602 (2001)
  [arXiv:hep-ph/0106295].
  %%CITATION = HEP-PH 0106295;%%
 \\
  G.~Landsberg,
  {\em ``Black holes at future colliders and in cosmic rays,''}
  Eur.\ Phys.\ J.\ C {\bf 33}, S927 (2004)
  [arXiv:hep-ex/0310034].
  %%CITATION = HEP-EX 0310034;%% 

%----------------------------------------------------------------  
\bibitem{others}
  R.~H.~Brandenberger and J.~Martin,
  {\em ``Back-reaction and the trans-Planckian problem of inflation revisited,''}
  Phys.\ Rev.\ D {\bf 71}, 023504 (2005)
  [arXiv:hep-th/0410223].
  %%CITATION = HEP-TH 0410223;%%
  
%----------------------------------------------------------------  
\bibitem{Rovelli}
  C.~Rovelli and S.~Speziale,
  {\em ``Reconcile Planck-scale discreteness and the Lorentz-Fitzgerald
  contraction,''}
  Phys.\ Rev.\ D {\bf 67}, 064019 (2003)
  [arXiv:gr-qc/0205108].
  %%CITATION = GR-QC 0205108;%%
  
%----------------------------------------------------------------
\bibitem{KS89}
V.~A.~Kostelecky and S.~Samuel, {\em ``Spontaneous Breaking Of Lorentz
Symmetry In String Theory,''} Phys.\ Rev.\ D {\bf 39}, 683 (1989).
%%CITATION = PHRVA,D39,683;%%

%---------------------------------
\bibitem{Damour:1994zq}
T.~Damour and A.~M.~Polyakov, {\em ``The String dilaton and a least
coupling principle,''} Nucl.\ Phys.\ B {\bf 423}, 532 (1994)
[arXiv:hep-th/9401069].
%%CITATION = HEP-TH 9401069;%%

%------------------------------
\bibitem{GAC-Nat}
  G.~Amelino-Camelia, J.~R.~Ellis, N.~E.~Mavromatos and D.~V.~Nanopoulos,
  {\em ``Distance measurement and wave dispersion in a Liouville-string approach  to
  quantum gravity,''}
  Int.\ J.\ Mod.\ Phys.\ A {\bf 12}, 607 (1997)
  [arXiv:hep-th/9605211];
  %%CITATION = HEP-TH 9605211;%%
\\
G.~Amelino-Camelia, J.~R.~Ellis, N.~E.~Mavromatos, D.~V.~Nanopoulos
and S.~Sarkar, {\em ``Potential Sensitivity of Gamma-Ray Burster
Observations to Wave Dispersion in Vacuo,''} Nature {\bf 393}, 763
(1998) [arXiv:astro-ph/9712103].
%%CITATION = ASTRO-PH 9712103;%%

%------------------------------
\bibitem{GP}
R.~Gambini and J.~Pullin, {\em ``Nonstandard optics from quantum
spacetime,''} Phys.\ Rev.\ D {\bf 59}, 124021 (1999)
[arXiv:gr-qc/9809038].
%%CITATION = GR-QC 9809038;%%

%----------------------------------------------------------------
\bibitem{loopqg}
J.~Alfaro, H.~A.~Morales-Tecotl and L.~F.~Urrutia, {\em ``Quantum gravity
corrections to neutrino propagation,''} Phys.\ Rev.\ Lett.\ {\bf 84},
2318 (2000) [arXiv:gr-qc/9909079];
%%CITATION = GR-QC 9909079;%%
\\
J.~Alfaro, H.~A.~Morales-Tecotl and L.~F.~Urrutia,{\em ``Loop quantum
gravity and light propagation,''} Phys.\ Rev.\ D {\bf 65}, 103509
(2002). [arXiv:hep-th/0108061].
%%CITATION = HEP-TH 0108061;%%

%---------------------------------
\bibitem{DSR}
G.~Amelino-Camelia, {\em ``Relativity in space-times with short-distance
structure governed by an observer-independent (Planckian) length
scale,''} Int.\ J.\ Mod.\ Phys.\ D {\bf 11}, 1643 (2002)
[arXiv:gr-qc/0210063].
%%CITATION = GR-QC 0210063;%%

%---------------------------------
\bibitem{DSRint}
  S.~Liberati, S.~Sonego and M.~Visser,
  {\em ``Interpreting doubly special relativity as a modified theory of
  measurement,''}
  Phys.\ Rev.\ D {\bf 71}, 045001 (2005)
  [arXiv:gr-qc/0410113].
  %%CITATION = GR-QC 0410113;%%
  
  
  \bibitem{Judes}
  S.~Judes and M.~Visser,
  ``Conservation Laws in Doubly Special Relativity,''
  Phys.\ Rev.\ D {\bf 68} (2003) 045001
  [arXiv:gr-qc/0205067].
  %%CITATION = GR-QC 0205067;%%
  


%----------------------------------------------------------------
\bibitem{Hayakawa}
M.~Hayakawa, {\em ``Perturbative analysis on infrared aspects of
noncommutative QED on  $R^4$,''} Phys.\ Lett.\ B {\bf 478}, 394
(2000) [arXiv:hep-th/9912094];
%%CITATION = HEP-TH 9912094;%%
\\
M.~Hayakawa, {\em ``Perturbative analysis on infrared and ultraviolet
aspects of noncommutative QED on R**4,''} [arXiv:hep-th/9912167].
%%CITATION = HEP-TH 9912167;%%

%----------------------------------------------------------------
\bibitem{Mocioiu:2000ip}
  I.~Mocioiu, M.~Pospelov and R.~Roiban,
 {\em ``Low-energy limits on the antisymmetric tensor field background on
the brane and on the non-commutative scale,''}
  Phys.\ Lett.\ B {\bf 489}, 390 (2000)
  [arXiv:hep-ph/0005191].
  %%CITATION = HEP-PH 0005191;%%

%----------------------------------------------------------------
\bibitem{Carroll:2001ws}
S.~M.~Carroll, J.~A.~Harvey, V.~A.~Kostelecky, C.~D.~Lane and
T.~Okamoto, {\em ``Noncommutative field theory and Lorentz violation,''}
Phys.\ Rev.\ Lett.\ {\bf 87}, 141601 (2001)
[arXiv:hep-th/0105082].
%%CITATION = HEP-TH 0105082;%%

%----------------------------------------------------------------
\bibitem{Anisimov:2001zc}
  A.~Anisimov, T.~Banks, M.~Dine and M.~Graesser,
  {\em ``Comments on non-commutative phenomenology,''}
  Phys.\ Rev.\ D {\bf 65}, 085032 (2002)
  [arXiv:hep-ph/0106356].
  %%CITATION = HEP-PH 0106356;%%

%----------------------------------------------------------------  
\bibitem{jlm-ann}
  T.~Jacobson, S.~Liberati and D.~Mattingly,
  {\em ``Lorentz violation at high energy: Concepts, phenomena and astrophysical
  constraints,''}
  Annals Phys.\  {\bf 321}, 150 (2006)
  [arXiv:astro-ph/0505267].
  %%CITATION = ASTRO-PH 0505267;%%

\bibitem{jlm-notes}
T.~Jacobson, S.~Liberati and D.~Mattingly,
  ``Astrophysical bounds on Planck suppressed Lorentz violation,''
  arXiv:hep-ph/0407370.
  %%CITATION = HEP-PH 0407370;%%

\bibitem{jlm-qgp}
T.~Jacobson, S.~Liberati and D.~Mattingly,
  ``Quantum gravity phenomenology and Lorentz violation,''
  arXiv:gr-qc/0404067.
  %%CITATION = GR-QC 0404067;%%

\bibitem{jlm-limits}
T.~A.~Jacobson, S.~Liberati, D.~Mattingly and F.~W.~Stecker,
  ``New limits on Planck scale Lorentz violation in QED,''
  Phys.\ Rev.\ Lett.\  {\bf 93} (2004) 021101
  [arXiv:astro-ph/0309681].
  %%CITATION = ASTRO-PH 0309681;%%
  
\bibitem{jlm-comments}
 T.~Jacobson, S.~Liberati and D.~Mattingly,
  ``Comments on 'Improved limit on quantum-spacetime modifications of  Lorentz
  symmetry from observations of gamma-ray blazars',''
  arXiv:gr-qc/0303001.
  %%CITATION = GR-QC 0303001;%% 
  
\bibitem{jlm-nature}
   T.~Jacobson, S.~Liberati and D.~Mattingly,
   ``A strong astrophysical constraint on the violation of special relativity by quantum gravity'', 
%  ``Lorentz violation and Crab synchrotron emission: A new constraint far
 % beyond the Planck scale,''
  Nature {\bf 424} (2003) 1019
  [arXiv:astro-ph/0212190].
  %%CITATION = ASTRO-PH 0212190;%%
  
  \bibitem{jlm-tev}
  T.~Jacobson, S.~Liberati and D.~Mattingly,
  ``TeV astrophysics constraints on Planck scale Lorentz violation,''
  Phys.\ Rev.\ D {\bf 66} (2002) 081302
  [arXiv:hep-ph/0112207].
  %%CITATION = HEP-PH 0112207;%%
 
  \bibitem{jlm-bloomington}
  S.~Liberati, T.~A.~Jacobson and D.~Mattingly,
  ``High energy constraints on Lorentz symmetry violations,''
  arXiv:hep-ph/0110094.
  %%CITATION = HEP-PH 0110094;%%
  
%------------------------------
\bibitem{GAC-crit}
G.~Amelino-Camelia, {\em ``Improved limit on quantum-spacetime
modifications of Lorentz symmetry from observations of gamma-ray
blazars,''} [arXiv:gr-qc/0212002];
%%CITATION = GR-QC 0212002;%%
\\
G.~Amelino-Camelia, 
{\em ``A perspective on quantum gravity phenomenology,''}
[arXiv:gr-qc/0402009].
%%CITATION = GR-QC 0402009;%%

%----------------------------------------------------------------    

\bibitem{ESM}
D.~Colladay and V.~A.~Kostelecky, {\em ``Lorentz-violating extension of
the standard model,''} Phys.\ Rev.\ D {\bf 58}, 116002 (1998),
[arXiv:hep-ph/9809521].
%%CITATION = HEP-PH 9809521;%%

%------------------------------
\bibitem{MP}
R.~C.~Myers and M.~Pospelov, 
{\em ``Experimental challenges for quantum
gravity,''} Phys.\ Rev.\ Lett.\  {\bf 90}, 211601 (2003)
[arXiv:hep-ph/0301124].
%%CITATION = HEP-PH 0301124;%%

%------------------------------
\bibitem{Collins}
  J.~Collins, A.~Perez, D.~Sudarsky, L.~Urrutia and H.~Vucetich,
{\em  ``Lorentz invariance: An additional fine-tuning problem,''}
  Phys.\ Rev.\ Lett.\  {\bf 93}, 191301 (2004)
  [arXiv:gr-qc/0403053].
  %%CITATION = GR-QC 0403053;%%

%------------------------------
\bibitem{Pospelov-Nibbelink}  
S.~G.~Nibbelink and M.~Pospelov, ``Lorentz violation in
supersymmetric field theories,'' Phys.\ Rev.\ Lett.\ {\bf 94},
081601 (2005) [arXiv:hep-ph/0404271];\\
%%CITATION = HEP-PH 0404271;%%
  P.~A.~Bolokhov, S.~G.~Nibbelink and M.~Pospelov,
 {\em ``Lorentz Violating Supersymmetric Quantum Electrodynamics,''}
  [arXiv:hep-ph/0505029].
  %%CITATION = HEP-PH 0505029;%%

%------------------------------
\bibitem{GAC-Pir}
G.~Amelino-Camelia and T.~Piran, ``Planck-scale deformation of
Lorentz symmetry as a solution to the UHECR and the TeV-gamma
paradoxes,'' Phys.\ Rev.\ D {\bf 64}, 036005 (2001)
[arXiv:astro-ph/0008107].
%%CITATION = ASTRO-PH 0008107;%%

%------------------------------
\bibitem{LivRev}
  C.~Barcel\'o, S.~Liberati and M.~Visser,
  {\em ``Analogue gravity,''}
  Living Rev. Rel. {\bf 8} (2005) 12 
 [arXiv:gr-qc/0505065].
  % Living Reviews in Relativity
  %%CITATION = GR-QC 0505065;%%
  
  \bibitem{Nielsen}
  C.~D.~Froggatt and H.~B.~Nielsen,
  ``Derivation of Poincare invariance from general quantum field theory,''
  arXiv:hep-th/0501149.
  %%CITATION = HEP-TH 0501149;%%
\\  
  C.~D.~Froggatt and H.~B.~Nielsen,
  ``Derivation of Lorentz invariance and three space dimensions in generic
  field theory,''
  arXiv:hep-ph/0211106.
  %%CITATION = HEP-PH 0211106;%%

\bibitem{Bjorken}
J.~D.~Bjorken,
  ``Emergent gauge bosons,''
  arXiv:hep-th/0111196.
  %%CITATION = HEP-TH 0111196;%%


\bibitem{Laughlin}
R.~B.~Laughlin,
  ``Emergent relativity,''
  Int.\ J.\ Mod.\ Phys.\ A {\bf 18} (2003) 831
  [arXiv:gr-qc/0302028].
  %%CITATION = GR-QC 0302028;%%


  
%------------------------------
\bibitem{BLV}
  C.~Barcel\'o, S.~Liberati and M.~Visser,
  {\em ``Analogue gravity from Bose-Einstein condensates,''}
  Class.\ Quant.\ Grav.\  {\bf 18}, 1137 (2001)
  [arXiv:gr-qc/0011026].
  %%CITATION = GR-QC 0011026;%%
  
%------------------------------  
\bibitem{Breakdown} 
M.~Visser, C.~Barcel\'o and S.~Liberati,
 {\em  ``Acoustics in Bose--Einstein condensates as an example of broken Lorentz
  symmetry,''}
  arXiv:hep-th/0109033.
  %%CITATION = HEP-TH 0109033;%%
  
%--------------------------------
\bibitem{Garay}
  L.~J.~Garay, J.~R.~Anglin, J.~I.~Cirac and P.~Zoller,
  {\em ``Black holes in Bose-Einstein condensates,''}
  Phys.\ Rev.\ Lett.\  {\bf 85}, 4643 (2000)
  [arXiv:gr-qc/0002015];\\
  %%CITATION = GR-QC 0002015;%%
  L.~J.~Garay, J.~R.~Anglin, J.~I.~Cirac and P.~Zoller,
  {\em ``Sonic black holes in dilute Bose-Einstein condensates,''}
  Phys.\ Rev.\ A {\bf 63}, 023611 (2001)
  [arXiv:gr-qc/0005131].
  %%CITATION = GR-QC 0005131;%%
%

\bibitem{VW1}
  M.~Visser and S.~Weinfurtner,
  {\em ``Massive Klein-Gordon equation from a BEC-based analogue spacetime,''}
  Phys.\ Rev.\ D {\bf 72} (2005) 044020
  [arXiv:gr-qc/0506029].
  %%CITATION = GR-QC 0506029;%%
  
\bibitem{VW2}
  M.~Visser and S.~Weinfurtner,
  {\em ``Massive phonon modes from a BEC-based analogue model,''}
  arXiv:cond-mat/0409639.
  %%CITATION = COND-MAT 0409639;%%
  
 
\bibitem{Fischer}
U.~R.~Fischer and R.~Schutzhold,
  ``Quantum simulation of cosmic inflation in two-component Bose-Einstein
  condensates,''
  Phys.\ Rev.\ A {\bf 70} (2004) 063615
  [arXiv:cond-mat/0406470].
  %%CITATION = COND-MAT 0406470;%%
  
\bibitem{abh}
M.~Visser,
  ``Acoustic black holes: Horizons, ergospheres, and Hawking radiation,''
  Class.\ Quant.\ Grav.\  {\bf 15} (1998) 1767
  [arXiv:gr-qc/9712010].
  %%CITATION = GR-QC 9712010;%%
 \\ 
M.~Visser,
  ``Acoustic propagation in fluids: An Unexpected example of Lorentzian
  geometry,''
  arXiv:gr-qc/9311028.
  %%CITATION = GR-QC 9311028;%%
\\
M.~Visser,
  ``Acoustic black holes,''
  arXiv:gr-qc/9901047.
  %%CITATION = GR-QC 9901047;%%
  \\
S.~Liberati, S.~Sonego and M.~Visser,
  ``Unexpectedly large surface gravities for acoustic horizons?,''
  Class.\ Quant.\ Grav.\  {\bf 17} (2000) 2903
  [arXiv:gr-qc/0003105].
  %%CITATION = GR-QC 0003105;%%
  \\
C.~Barcel\'o, S.~Liberati, S.~Sonego and M.~Visser,
  ``Causal structure of analogue spacetimes,''
  New J.\ Phys.\  {\bf 6} (2004) 186
  [arXiv:gr-qc/0408022].
  %%CITATION = GR-QC 0408022;%%
  \\
M.~Visser,
  ``Hawking radiation without black hole entropy,''
  Phys.\ Rev.\ Lett.\  {\bf 80} (1998) 3436
  [arXiv:gr-qc/9712016].
  %%CITATION = GR-QC 9712016;%%
  \\
  C.~Barcel\'o, S.~Liberati and M.~Visser,
  ``Towards the observation of Hawking radiation in Bose-Einstein
  condensates,''
  Int.\ J.\ Mod.\ Phys.\ A {\bf 18} (2003) 3735
  [arXiv:gr-qc/0110036].
  %%CITATION = GR-QC 0110036;%%
  \\
  C.~Barcel\'o, S.~Liberati and M.~Visser,
  ``Probing semiclassical analogue gravity in Bose--Einstein condensates with
  widely tunable interactions,''
  Phys.\ Rev.\ A {\bf 68} (2003) 053613
  [arXiv:cond-mat/0307491].
  %%CITATION = COND-MAT 0307491;%%
  \\
 M.~Visser, C.~Barcel\'o and S.~Liberati,
  ``Analogue models of and for gravity,''
  Gen.\ Rel.\ Grav.\  {\bf 34} (2002) 1719
  [arXiv:gr-qc/0111111].
  %%CITATION = GR-QC 0111111;%%
 \\
C.~Barcel\'o, S.~Liberati and M.~Visser,
  ``Analogue models for FRW cosmologies,''
  Int.\ J.\ Mod.\ Phys.\ D {\bf 12} (2003) 1641
  [arXiv:gr-qc/0305061].
  %%CITATION = GR-QC 0305061;%%

\bibitem{silke}
S.~E.~C.~Weinfurtner,
  ``Analogue model for an expanding universe,''
  arXiv:gr-qc/0404063.
  %%CITATION = GR-QC 0404063;%%
\\
S.~E.~C.~Weinfurtner,
  ``Simulation of gravitational objects in Bose-Einstein condensates,''
  arXiv:gr-qc/0404022.
  %%CITATION = GR-QC 0404022;%%
  
\bibitem{Schutzhold:cosmic}
R.~Schutzhold,
  ``Dynamical zero-temperature phase transitions and cosmic inflation /
  deflation,''
  Phys.\ Rev.\ Lett.\  {\bf 95} (2005) 135703
  [arXiv:quant-ph/0505196].
  %%CITATION = QUANT-PH 0505196;%%

  %2-comp. BEC:

\bibitem{trippenbach}
M.~Trippenbach, K.~G$\acute{o}$ral, K.~Rza$\dot{z}$ewski, B.~Malomed,
and Y.~B.~Band, {\em ``Structure of binary Bose--Einstein
condensates,''} J.\ Phys.\ B {\bf 33}, 4017 (2000)
 [arXiv:cond-mat/0008255];

\bibitem{jenkins}
S.~D.~Jenkins and T.~A.~B.~Kennedy,
  {\em ``Dynamic stability of dressed condensate mixtures,''}
  Phys.\ Rev.\ A {\bf 68}, 053607 (2003);
  
\bibitem{Toms}
D.~J.~Toms  
{\em $\zeta$-function regularization and the interacting Bose gas at low temperature}
  Phys.\ Rev.\ A {\bf 66}, 013619 (2002);
  
 \bibitem{abh:emergent}
 C.~Barcel\'o, M.~Visser and S.~Liberati,
  ``Einstein gravity as an emergent phenomenon?,''
  Int.\ J.\ Mod.\ Phys.\ D {\bf 10} (2001) 799
  [arXiv:gr-qc/0106002].
  %%CITATION = GR-QC 0106002;%%



\bibitem{normal}
C.~Barcel\'o, S.~Liberati and M.~Visser,
  ``Analogue gravity from field theory normal modes?,''
  Class.\ Quant.\ Grav.\  {\bf 18} (2001) 3595
  [arXiv:gr-qc/0104001].
  %%CITATION = GR-QC 0104001;%%

\bibitem{refringence}
  C.~Barcel\'o, S.~Liberati and M.~Visser,
  ``Refringence, field theory, and normal modes,''
  Class.\ Quant.\ Grav.\  {\bf 19}, 2961 (2002)
  [arXiv:gr-qc/0111059].
  %%CITATION = GR-QC 0111059;%%

\bibitem{bimetricity}
M.~Visser, C.~Barcel\'o and S.~Liberati,
  ``Bi-refringence versus bi-metricity,''
  arXiv:gr-qc/0204017.
  %%CITATION = GR-QC 0204017;%%

\bibitem{Vogels:2002qd}
  J.~M.~Vogels, K.~Xu, C.~Raman, J.~R.~Abo-Shaeer and W.~Ketterle,
  %``Experimental observation of the Bogoliubov transformation for a
  %Bose-Einstein condensed gas,''
  Phys.\ Rev.\ Lett.\  {\bf 88} (2002) 060402
  [arXiv:cond-mat/0109205].
  %%CITATION = COND-MAT 0109205;%%


%%uwe and ralf:

%\bibitem{Schutzhold:2005ex}
%  R.~Schutzhold, M.~Uhlmann, Y.~Xu and U.~R.~Fischer,
%  ``Quantum back-reaction in dilute Bose-Einstein condensates,''
%  Phys.\ Rev.\ D {\bf 72} (2005) 105005
%  [arXiv:cond-mat/0503581].
%  %%CITATION = COND-MAT 0503581;%%
  
\bibitem{Luty}
  H.~S.~Goh, M.~A.~Luty and S.~P.~Ng,
  %``Supersymmetry without supersymmetry,''
  JHEP {\bf 0501}, 040 (2005)
  [arXiv:hep-th/0309103].
  %%CITATION = HEP-TH 0309103;%%
%%%%%%%%%%%%%%%%%%%%%%%%%%%%%%%%%%%%%%%%%%%
%%%%%%%%%%%%%%%%%%%%%%%%%%%%%%%%%%%%%%%%%%%%


%-----------------------------------
\end{thebibliography}
\end{document}